# Bachelier's Market Model for ESG Asset Pricing

Svetlozar Rachev[1], Nancy Asare Nyarko[2], Blessing Omotade[3], and Peter Yegon[4]

This paper is dedicated to my dear friend and colleague Professor Ashis SenGupta on the occasion of his 70th birthday. I still remember the great time we had together writing the paper on the use of the Laplace-Weibull mixture in modeling asset returns for *Management Science* in 1993. That was the starting point of my work on rewriting the asset pricing theory for non-Gaussian, heavy-tailed distributions. Thank you, Ashis. I wish you many happy years.

## Abstract

Environmental, Social, and Governance (ESG) finance is a cornerstone of modern finance and investment, as it changes the classical return-risk view of investment by incorporating an additional dimension of investment performance: the ESG score of the investment. We define the ESG price process and integrate it into an extension of Bachelier's market model in both discrete and continuous time, enabling option pricing valuation.

[1] Svetlozar (Zari) Rachev, Department of Mathematics & Statistics, Texas Tech University, Box 41042, Lubbock, TX 79409-1042 USA, email: zari.rachev@ttu.edu

[2] Department of Mathematics & Statistics, Texas Tech University, Box 41042, Lubbock, TX 79409-1042 USA, email: nasareny@ttu.edu

[3] Department of Mathematics & Statistics, Texas Tech University, Box 41042, Lubbock, TX 79409-1042 USA, email: bomotade@ttu.edu

[4] Department of Mathematics & Statistics, Texas Tech University, Box 41042, Lubbock, TX 79409-1042 USA, email: pyegon@ttu.edu

# 1. Introduction

In his Ph.D. thesis [1], Louis Bachelier pioneered an option pricing model, which marked the birth of mathematical finance [2–4]. The price dynamics in Bachelier's model follow arithmetic Brownian motion, and although it was initially rejected as a suitable model for asset prices, there have been a few exceptions. Bachelier's market model necessitates the introduction of a riskless bank account with a simpler riskless interest rate. This extends Shiryaev's [5] exposition of Bachelier's market model with zero interest rates. We argue that combining Bachelier's risky asset dynamics of arithmetic Brownian motion with a riskless bank account using compound interest rates, as seen in [6] and [7], is not appropriate. The criticism that Bachelier's market model allows negative riskless rates is now irrelevant, as real financial markets have demonstrated the existence of such cases [8].

Currently, the main challenge in using Bachelier's market model is that asset prices must be non-negative. While stock prices are indeed non-negative, the ESG-adjusted price can be negative. In this paper, we define **the ESG-adjusted stock price** $S_t^{ESG}, t \geq 0$, as follows:

$$S_t^{ESG} = S_t^{(X)}\left(1 + \gamma^{ESG} Z_t^{(X;I)}\right) \in R,$$

where $S_t^{(X)} > 0$ is the stock price of the company $X$, $Z_t^{(X;I)} \in R$ is **the relative ESG score** of $X$, and $\gamma^{ESG} \in R$ is the **ESG affinity of the financial market**. The ESG-adjusted stock price possesses the desirable characteristic of introducing a third dimension (ESG) into the traditional two-dimensional (risk-return) investment process, as discussed by Lauria et al. [9] and Hu et al. [10]. Some financial industry specialists and financial regulators currently consider the ESG performance as an independent factor, alongside the risk and return, which are the other two factors [11].

ESG-conscious option traders may choose to incorporate the impact of ESG factors on option valuation, potentially resulting in lower valuations for options for stocks with a low ESG affinity. They can calculate the ESG fair value of their option contracts using option valuation methods based on Bachelier's pricing formulas, as proposed in our suggested models. Following the approach of Hu et al. [12–14], our discrete Bachelier's option pricing approach maintains the parameters of the spot ESG prices. As a result, these spot market parameters of Bachelier's market model can be calibrated using existing option market prices. A comprehensive calibration analysis will be conducted in our future work. The paper is organized as follows. In Section 2, we present the Bachelier market model in continuous time, building upon Shiryaev's [5] exposition of Bachelier's market model with zero interest rates. We introduce a simple interest account as the riskless asset in Bachelier's market model, motivated by the incorporation of the ESG-adjusted stock price as the risky asset. Section 3 offers a concise overview of the methodology outlined in Hu et al. [12–14] for discrete option pricing, which preserves the spot parameters of the underlying asset. We apply this approach in our Bachelier's option pricing models. In Section 4, we derive the binomial option pricing in Bachelier's market model for independent price changes. By utilizing the Donsker-Prokhorov invariance principle, we then transition to continuous-time Bachelier's option pricing. Section 5 extends the results from Section 4 to encompass dependent price changes. Here, we rely on the Cherny-Shiryaev-Yor invariance principle as the primary tool. Appendix A introduces the ESG price as a natural extension of the observed stock price in the market. We discuss the suitability of Bachelier's market model when dealing with the ESG score as an independent factor in the risk-return profiling of stocks. Appendix B introduces the generalized skew Brownian motion as an uncertainty driver in Bachelier's market model. Finally, the concluding remarks are presented in Section 6, which summarizes the key findings and implications of our study.

## 2. Bachelier option pricing formula in a continuous-time market with a riskless simple interest account

The option contract considered by Louis Bachelier was a contract on exchange-traded futures over French government rentes [15]. Under the Bachelier model, the forward swap rate [16] $F_t^{(swap)}, t \in [0, T]$, has the price dynamics of arithmetic Brownian motion with zero drift:

$$F^{(swap)}(t) = F^{(swap)}(0) + \sigma^{(swap)} B_t, t \in [0, T],$$

where $\sigma^{(swap)} > 0$ is the Bachelier swap rate volatility and $B_t, t \in [0, T]$, is a standard Brownian motion[5]. The Bachelier formulas for the call option (payer swaption) and the put option (receiver swaption) are given in [16].

Following Shiryaev's [5] exposition of Bachelier's market model, the risky asset $\mathcal{A}^{(0)}$ has the price dynamics of an arithmetic Brownian motion

$$(2.1) \qquad A_t^{(0)} = A_0 + \rho t + v B_t, t \in [0, T], \rho > 0, v > 0, A_0 > 0,$$

where $B_t, t \in [0, T]$, is a standard Brownian motion on a stochastic basis (filtered probability space) $(\Omega, \mathcal{F}, \mathbb{F} = \{\mathcal{F}_t = \sigma(B_u, u \le t) \subseteq \mathcal{F}, t \in [0, T]\}, \mathbb{P})$ on a complete probability space $(\Omega, \mathcal{F}, \mathbb{P})$. The riskless asset $\mathcal{B}_0$ in Bachelier's market model is a riskless bank account $\beta_t^{(0)}, t \in [0, T]$, with a riskless rate $r_0 = 0$, that is, $\beta_t^{(0)} = 1, t \in [0, T]$. In this setting, Bachelier's market model $(\mathcal{A}^{(0)}, \mathfrak{B}^{(0)})$ is arbitrage-free and complete. The unique equivalent martingale measure $\mathbb{Q}^{(0)} \sim \mathbb{P}$ is determined by $d\mathbb{Q}^{(0)} = Z_T^{(0)} d\mathbb{P}$, where the Radon-Nikodym derivative $Z_T$ is given by

$$(2.2) \qquad Z_T^{(0)} = \exp\left(-\theta_0 B_T - \frac{1}{2}\theta_0^2 T\right),$$

where $\theta_0 = \frac{\rho}{v} > 0$ is the market price of risk. The Brownian motion $B_t^{\mathbb{Q}^{(0)}}$ on $(\Omega, \mathcal{F}, \mathbb{F}, \mathbb{Q}^{(0)})$ is defined as an arithmetic Brownian motion $B_t^{\mathbb{Q}^{(0)}} = B_t + \theta_0 t$ on $\mathbb{P}$. On $\mathbb{Q}^{(0)}$, the price dynamics of $\mathcal{A}^{(0)}$ are given by "Brownian motion process: The most important stochastic process is the Brownian motion or Wiener process. It was first discussed by Louis Bachelier (1900), who was interested in modeling fluctuations in prices in financial markets, and by Albert Einstein (1905), who gave a mathematical model for the irregular motion of colloidal particles first observed by the Scottish botanist Robert Brown in 1827. The first mathematically rigorous treatment of this model was given by Wiener (1923)."

---

[5]From Britannica (https://www.britannica.com/science/probability-theory/Brownian-motion-process#ref407454):

$$(2.3) \qquad A_t^{(0)} = A_0 + \nu B_t^{\mathbb{Q}^{(0)}}.$$

Shiryaev's [5] exposition considers a European call option contact $\mathcal{C}^{(0)}$ with a strike $K$ and a terminal time $T > 0$, that is, $C_T^{(0)} = \max(A_T^{(0)} - K, 0)$. Then, the call price $C_t^{(0)}, t \in [0, T]$, will be given by $C_t^{(0)} = \mathbb{E}_t^{\mathbb{Q}^{(0)}}\left(C_T^{(0)}\right)$. This risk-neutral valuation of $C_t^{(0)}$ leads to Bachelier's formula for $\mathcal{C}^{(0)}$:

$$(2.4) \qquad C_t^{(0)} = \left(A_t^{(0)} - K\right)\Phi\left(\frac{A_t^{(0)} - K}{\nu\sqrt{T-t}}\right) + \nu\sqrt{T-t}\,\varphi\left(\frac{A_t^{(0)} - K}{\nu\sqrt{T-t}}\right),$$

where $\Phi$ is the cumulative probability distribution of a standard normal random variable, and $\varphi = \Phi'$ is its probability density [5, pp. 737–738].

The hedging strategy is determined as in the classical Black-Scholes-Merton option pricing formula [17, Section 5F]. From (2.1) and assuming that $C_t^{(0)} = C^{(0)}\left(A_t^{(0)}, t\right), t \in [0, T)$, where $C^{(0)}(x, t), x \in R = (-\infty, \infty), t \in \{0, T\}$, has continuous derivatives $\frac{\partial C^{(0)}(x,t)}{\partial t}, \frac{\partial^2 C^{(0)}(x,t)}{\partial x^2}, x \in R, t \in \{0, T\}$, we have that $C_t^{(0)} = C^{(0)}\left(A_t^{(0)}, t\right)$ is an Itô process, with

$$(2.5) \qquad dC_t^{(0)} = \left(\frac{\partial C^{(0)}\left(A_t^{(0)}, t\right)}{\partial t} + \frac{\partial C\left(A_t^{(0)}, t\right)}{\partial x}\rho + \frac{1}{2}\frac{\partial^2 C^{(0)}\left(A_t^{(0)}, t\right)}{\partial x^2}\nu^2\right)dt + \frac{\partial C^{(0)}\left(A_t^{(0)}, t\right)}{\partial x}\nu dB_t.$$

Under the no-arbitrage assumptions for the market $\left(\mathcal{A}, \mathfrak{B}^{(0)}, \mathcal{C}\right)$, the hedger, taking the short position in $\mathcal{C}$, must form a self-financing replicating portfolio $P_t^{(0)} = a_t A_t^{(0)} + b_t \beta_t^{(0)} = C_t^{(0)}$. With $\beta_t^{(0)} = 1$, the standard no-arbitrage arguments and the stochastic differential equation (SDE) (2.5) lead to

$$(2.6) \qquad a_t = \frac{\partial C^{(0)}\left(A_t^{(0)}, t\right)}{\partial x}, b_t = C^{(0)}\left(A_t^{(0)}, t\right) - \frac{\partial C^{(0)}\left(A_t^{(0)}, t\right)}{\partial x}A_t^{(0)},$$

and, consequently, to Bachelier's partial differential equation (PDE):

$$(2.7) \qquad \frac{\partial C^{(0)}(x,t)}{\partial t} + \frac{1}{2}\frac{\partial^2 C^{(0)}(x,t)}{\partial x^2}\nu^2 = 0.$$

The self-financing strategy $(a_t, b_t)$ in (2.6) has an explicit form [5, p. 738]:

$$(2.8) \qquad a_t = \Phi\left(\frac{A_t^{(0)} - K}{v\sqrt{T-t}}\right), b_t = -K\Phi\left(\frac{A_t^{(0)} - K}{v\sqrt{T-t}}\right) + v\sqrt{T-t}\,\varphi\left(\frac{A_t^{(0)} - K}{v\sqrt{T-t}}\right).$$

The restriction on the riskless rate, $r_0 = 0$, was based on the presumption that interest rates should be always non-negative. However, this is no longer always true in real-world financial practice [18–25]. Central banks have implemented negative interest rates in various countries and regions at different times. Some of the countries with negative interest rates as of 2023 are listed in Table 1 [8].

**Table 1** Countries with negative interest rates as of 2023

| Country | Interest Rate |
|---|---|
| Switzerland | -0.08% |
| Denmark | -0.06% |
| Japan | -0.01% |

We assume that on $(\Omega, \mathcal{F}, \mathbb{F}, \mathbb{P})$, the risky asset $\mathcal{A}$ has the price dynamics of an arithmetic Brownian motion

$$(2.9) \qquad A_t = A_0 + \rho t + v B_t, t \in [0, T], A_0 > 0, \rho \in R, v > 0.$$

As we will argue in Appendix A, the main motivation for deciding that the risky asset $\mathcal{A}$ will have the dynamics described by (2.9) is the ESG-adjusted stock price:

$$A_t = S_t^{ESG} = S_t^{(X)}\left(1 + \gamma^{ESG} Z_t^{(X;I)}\right) \in R,$$

where $S_t^{(X)} > 0$ is the stock price of the company $X$, $Z_t^{(X;I)} \in R$ is the relative ESG score of $X$, and $\gamma^{ESG} \in R$ is the **ESG affinity of the financial market[6]**. In this setting, the asset $\mathcal{A}$ is the ESG-adjusted stock.

Additional examples of price dynamics with positive and negative values are the spread of two stock prices, the spread between the stock price and a tradable benchmark security (stock index, US Treasury securities, etc.), the spreads of currency pairs, future and option contracts, and the yield differentials of stocks or bonds [26, 27]. Commodity spreads can be the difference between the prices of different commodities, the difference between the prices of the same commodity when it is traded in different locations, or the spreads of a commodity's prices at different moments in the production process, such as crack spreads in the oil market, crush spreads in the soybean market, spark spreads in the electricity market, calendar spreads of the prices of one commodity at different times, etc. [28].

Next, we modify Bachelier's market model $\left(\mathcal{A}, \mathfrak{B}^{(0)}, \mathcal{C}\right)$ by replacing the riskless bank account $\mathfrak{B}^{(0)}$ with the simple interest account (SIA) $\mathfrak{B}^{(SIA)}$, which we will define in the following.

The **simple interest account (SIA) $\mathfrak{B}^{(SIA)}$** with the **riskless simple interest rate $r^{(si)} \in R$** should have the same linear price dynamics as $A_t$ in (2.9) but with a volatility $v = 0$. We define the price process dynamics for $\mathfrak{B}^{(SIA)}$ as follows:

(2.10) $$\beta_t = \beta_0 + r^{(si)}t, t \in [0,T], \beta_0 > 0, r^{(si)} \in (-\infty, \rho).$$

Formally, $r^{(si)}$ represents a simple interest rate on a deposit of one unit of currency, as defined in [29]. Indeed, $\mathfrak{B}^{(SIA)} \equiv \mathcal{B}_0$ if and only if $r^{(si)} = r = 0$[7].

---

[6] The affinity of the financial market, $\gamma^{ESG}$, will be time-dependent in reality, but it will change at a much slower pace than market prices. However, for the considered period of time, $t \in [0,T]$, we prefer to view $\gamma^{ESG}$ as constant.

[7] Our exposition of Bachelier's market model can be easily extended to asset price dynamics following continuous diffusion, which is done in Chapters 6 and 7 of [17] for the Black-Scholes-

**Remark 1.** *In the US, the three-month US Treasury bill rate is considered a proxy for the short-term riskless rate [30]. The long-term average of the three-month US Treasury bill rate is 4.17%[8]. In Figure 1a, we show the relative difference $E(t) = \frac{\beta_t - b(t)}{b(t)}$ between the riskless bank account $b(t) = e^{r_{0.25}t}, t \in [0, \frac{1}{4}], r_{0.25} = 0.0417$, and the simple interest rate account with the same rate, $\beta_t = 1 + r_{0.25}t, t \in \left[0, \frac{1}{4}\right]$, over a three-month period. The relative difference is negligible.*

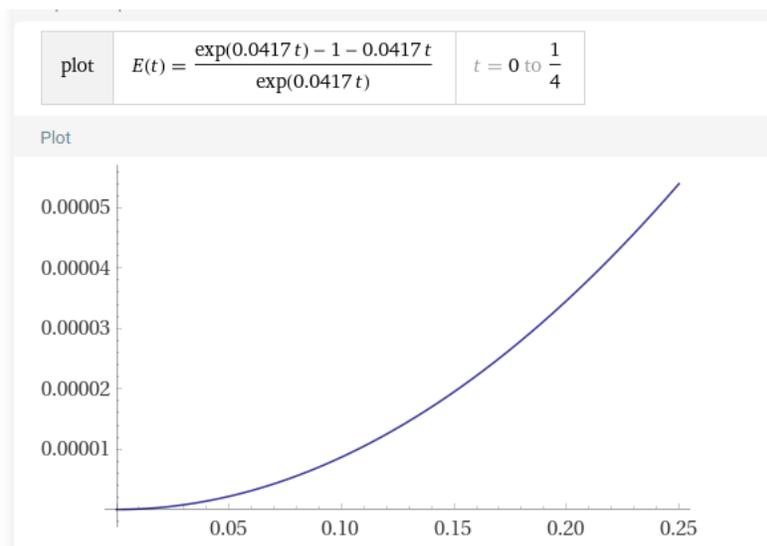

Merton market model, by replacing (2.9) with $dA_t = \rho_t dt + v_t dB_t, A_0 > 0, \rho_t \in R, \ t \in [0, T]$, and (2.10) with $d\beta_t = r_t^{(si)} dt, t \in [0, T], \beta_0 > 0, r_t^{(si)} \in (-\infty, \rho_t)$.

**Fig. 1a  The relative difference between the three-month Treasury bill rate and the simple interest over a three-month period. The three-month Treasury bill rate is chosen to be the long-term average of 4.17%.**

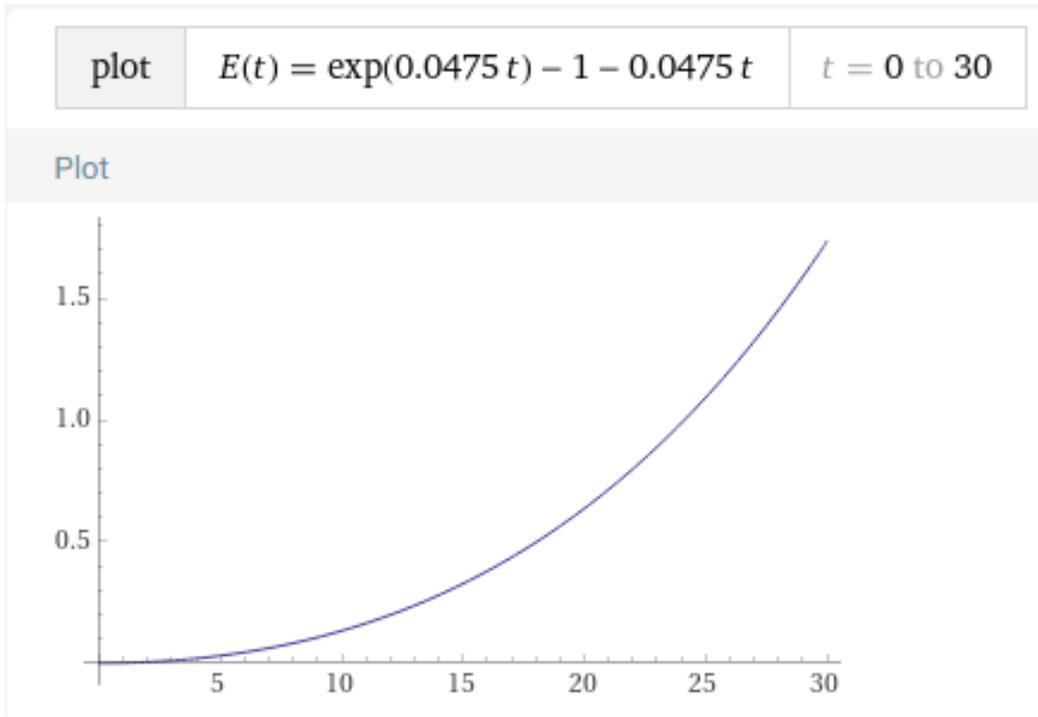

**Fig.  1b The relative difference between the 30-year Treasury rate and the simple interest over a 30-year period. The 30-year Treasury rate is chosen to be the long-term average of 4.74%.**

*As shown in Figure 1b, there is a significant relative difference when the 30-year Treasury rate[9] and the simple interest rate account, both with a rate of 0.0475, are compared. We argue*

---

*that in Bachelier's market model, the riskless asset should be the simple interest rate account, and it should not be replaced by a riskless bank account, as in the Black-Scholes-Merton market model. This is due to the consistency in the definitions of the price dynamics for the risky and riskless assets, as defined in (2.1) and (2.10). We shall provide more discussion of the choice of $\mathfrak{B}^{(SIA)}$ as the riskless asset in Bachelier's market model $\left(\mathcal{A}, \mathfrak{B}^{(SIA)}, \mathcal{C}\right)$ later in this section.*

Continuing our price-structure analysis of Bachelier's market model $\left(\mathcal{A}, \mathfrak{B}^{(SIA)}, \mathcal{C}\right)$, we search for the unique equivalent martingale measure $\mathbb{Q} \sim \mathbb{P}$ such that on $\mathbb{Q}$,

(2.11) $$dA_t = r^{(si)}dt + v dB_t^{(\mathbb{Q})}, \ \ t \in [0, T],$$

where $B_t^{(\mathbb{Q})}$ is a standard Brownian motion on $\mathbb{Q}$. On $\mathbb{P}$, $B_t^{(\mathbb{Q})}$ is an arithmetic Brownian motion $B_t^{(\mathbb{Q})} = B_t + \theta t$. Then, choosing $\theta = \frac{\rho - r^{(si)}}{v} > 0$ to be the market price of risk, we have $dA_t = r^{(si)}dt + v dB_t^{(\mathbb{Q})}$ on $\mathbb{Q}$. Thus, the *excess asset price over the **riskless simple interest rate**,*

$$A_t - \beta_t = A_0 - \beta_0 + v B_t^{(\mathbb{Q})},$$

is a martingale on $(\Omega, \mathcal{F}, \mathbb{F}, \mathbb{Q})$. This is the main difference between Bachelier's market model and the Black-Scholes-Merton market model, where the discounted price process under the risk-neutral measure is a martingale.

Consider a European call option contract $\mathcal{C}$ with a strike $K$ and a terminal time $T > 0$, that is, $C_T = \max(A_T - K, 0)$. Then, the call option price in excess of the riskless simple interest rate is a martingale under $\mathbb{Q}$, and

$$C_t - \beta_t = E_t^{\mathbb{Q}}(C_T - \beta_T) = E_t^{\mathbb{Q}}\left(\max\left[A_t + r^{(si)}(T-t) + v\left(B_T^{\mathbb{Q}} - B_t^{\mathbb{Q}}\right) - K, 0\right] - \beta_T\right).$$

---

important when looking at the overall US economy. Historically, the 30 year treasury yield reached upwards of 15.21% in 1981 when the Federal Reserve raised benchmark rates to contain inflation. The 30 Year yield also went as low as 2% in the low rate environment after the Great Recession."

Thus, $C_t = \mathbb{E}\left(\max\left[\left(A_t - K + r^{(si)}(T-t)\right) + v\sqrt{T-t}\mathbb{N}^{[0,1]}, 0\right]\right) - r^{(si)}(T-t)$, where $\mathbb{N}^{[0,1]}$ is a standard normal random variable. Then, $C_t = \mathbb{E}\left(\max\left[\lambda_{0,t} + \lambda_{1,t}\mathbb{N}^{[0,1]}, 0\right]\right) - r^{(si)}(T-t)$, where $\lambda_{0,t} = A_t - K + r^{(si)}(T-t) \in R$ and $\lambda_{1,t} = v\sqrt{T-t} \geq 0$. Finally, as in [5, p. 737], we have $\mathbb{E}\left(\max\left[\lambda_{0,t} + \lambda_{1,t}\mathbb{N}^{[0,1]}, 0\right]\right) = \lambda_{0,t}\Phi\left(\frac{\lambda_{0,t}}{\lambda_{1,t}}\right) + \lambda_{1,t}\varphi\left(\frac{\lambda_{0,t}}{\lambda_{1,t}}\right)$.

In summary, we have the following basic theorem.

**Theorem 1 (Bachelier's call option pricing in market with riskless simple interest rate).** *The call price in Bachelier's market model $\left(\mathcal{A}, \mathfrak{B}^{(SIA)}, \mathcal{C}\right)$ is given by*

(2.12)     $C_t = \lambda_{0,t}\Phi\left(\frac{\lambda_{0,t}}{\lambda_{1,t}}\right) + \lambda_{1,t}\varphi\left(\frac{\lambda_{0,t}}{\lambda_{1,t}}\right) - r^{(si)}(T-t), t \in [0,T],$

*where $\lambda_{0,t} = A_t - K + r^{(si)}(T-t)$, $\lambda_{1,t} = v\sqrt{T-t}$, $\Phi$ is the cumulative probability distribution of a standard normal random variable, and $\varphi = \Phi'$ is its probability density.*

Let us return to the necessity of choosing $\mathfrak{B}^{(SIA)}$ as the riskless asset when the risky asset $\mathcal{A}$ has the price dynamics described by (2.9). Consider a no-arbitrage complete market with only two risky assets that have the price dynamics described by (2.9) and no risk-free asset. The market's completeness requires that Bachelier's price dynamics for both assets must be determined by a single Brownian motion:

(2.13)        $dA_t^{(i)} = \rho_i dt + v_i dB_t, t \geq 0, A_0^{(i)} > 0, \rho_i \in R, v_i > 0, \ v_1 \neq v_2, \ i = 1,2.$

Consequently, the market prices of risk for both assets must be equal, that is, $\frac{\rho_1 - r^{(si)}}{v_1} = \frac{\rho_2 - r^{(si)}}{v_2}$ and thus the riskless rate is uniquely determined by

(2.14)        $r^{(si)} = \frac{\rho_1 v_2 - \rho_2 v_1}{v_2 - v_1}.$

The case of $(n+1)$ risky Bachelier's assets with dynamics determined by the $n$-dimensional Brownian motion $\mathbb{B}_t = \left(B_t^{(1)}, \ldots, B_t^{(n)}\right), t \geq 0$, which can be described by

(2.15)        $dA_t^{(i)} = \rho_i dt + \sum_{j=1}^{n} v_{i,j} dB_t^{(j)}, t \geq 0, i = 1, \ldots, n+1,$

is considered in a fashion similar to the fashion in which it is considered in [31], with the market price of risk defined as in [17, p. 113].

In [6], the stock price dynamics in Bachelier's market model are defined by the Ornstein-Uhlenbeck process

(2.16) $$dS_t = \mu S_t dt + \sigma dB_t, t \geq 0,$$

and thus $\mu \leq 0$ [32, p. 358]. The riskless bond dynamics are modeled by

$dP(t, T) = rP(t, T)dt, 0 \leq t < T, P(T, T) = 1$, where $r$ is the riskless rate. For the market price for risk to be positive, $r < \mu \leq 0$ must hold.

Brooks and Brooks [7] considered a more general price model

(2.17) $$dS_t = \mu(S_t, t)dt + \sigma(t)dB_t, t \geq 0,$$

on $(\Omega, \mathbb{F}, \mathbb{P})$. From (2.17), the dynamics of the riskless bank account should be determined by

$d\mathcal{b}_t = r(t)dt, t \geq 0$, in order to be consistent with the risky asset dynamics given by (2.17). The equivalent martingale measure $\mathbb{Q}$ is determined by the Brownian motion $B_t^{(\mathbb{Q})}, t \geq 0$, with $dB_t^{(\mathbb{Q})} = dB_t + \theta_t dt$ on $\mathbb{P}$. The market price of risk $\theta_t, t \geq 0$, is $\theta_t = \frac{\mu(S_t, t) - r(t)}{\sigma(t)}, t \geq 0$. Next, as in (2.14), we should be able to recover the risk-free asset dynamics. Suppose that we have a market with only two risky assets that have the following price dynamics: $dS_t^{(i)} = \mu\left(S_t^{(i)}, t\right)dt + \sigma_i(t)dB_t, i = 1,2, \sigma_1(t) \neq \sigma_2(t), t \geq 0$. Then, the riskless rate should be determined by considering the equality of the market prices of risk for both assets: $\frac{\mu\left(S_t^{(1)}, t\right) - r(t)}{\sigma_1(t)} = \frac{\mu\left(S_t^{(2)}, t\right) - r(t)}{\sigma_2(t)}$. This leads to

$$r(t) = \frac{\mu\left(S_t^{(1)}, t\right)\sigma_2(t) - \mu\left(S_t^{(2)}, t\right)\sigma_1(t)}{\sigma_2(t) - \sigma_1(t)},$$

and thus the riskless rate depends on the price levels $S_t^{(i)}, i = 1,2$. This is in contrast to the case with two geometric Brownian motions,

$$dS_t^{(i)} = \mu_t^{(i)}S_t^{(i)}dt + \sigma_t^{(i)}S_t^{(i)}dB_t, t \geq 0, i = 1,2, \sigma_t^{(1)} \neq \sigma_t^{(2)},$$

where the riskless rate

$$r_t = \frac{\mu_t^{(1)}\sigma_t^{(2)} - \mu_t^{(2)}\sigma_t^{(1)}}{\sigma_t^{(2)} - \sigma_t^{(1)}}$$

is independent of the price levels $S_t^{(i)}, i = 1,2$; this seems to be a more realistic way to model the riskless rate. Similarly, by considering Bachelier's price processes with time-dependent parameters, which can be described by

(2.18)     $$dA_t^{(i)} = \rho_i(t)dt + v_i(t)dB_t, t \geq 0, A_0^{(i)} \in R, \rho_i \in R, v_i > 0, \ i = 1,2,$$

we find that the riskless simple interest rate

(2.19)     $$r_t^{(si)} = \frac{\rho_1(t)v_2(t) - \rho_2(t)v_1(t)}{v_2(t) - v_1(t)}$$

is independent of the asset price level $A_t^{(i)}, i = 1,2$. This motivated us to model the Bachelier asset price dynamics as a continuous diffusion process (arithmetic Brownian motion with time-dependent parameters):

(2.20)     $$dA_t = \rho_t dt + v_t B_t, t \in [0,T], A_0 \in R, \rho_t \in R, v_t > 0.$$

According to (2.20), the corresponding price process for $\mathfrak{B}^{(SIA)}$ is

(2.21)     $$d\beta_t = r_t^{(si)}dt, t \in [0,T], \beta_0 > 0, r_t^{(si)} \in (-\infty, \rho_t),$$

where $r_t^{(si)}$ is the time-dependent riskless simple interest rate.

Melnikov and Wan [33] considered a modification of Bachelier's model with absorption. The price process $S_t^{(MW)}, \ t \in [0,T]$, is given by

(2.22)     $$S_{t_\tau}^{(MW)} = S_0 + \mu(t_\tau) + \sigma B_{t_\tau}, S_0 > 0, t \in [0,T],$$

where $t_\tau = \min(t, \tau)$ and $\tau = \inf\left\{t : \in [0, T], S_t^{(MW)} = 0\right\}$. The price process $S_t^{(MW)}$, $t \in [0, T]$, is non-negative, and when it hits zero, it stays there until time $T$. In the setting described by (2.20), $A_t, t \geq 0$, can be interpreted as the ESG-adjusted stock price

$$(2.23) \qquad A_t = S_t^{ESG} = S_t^{(X)}\left(1 + \gamma^{ESG} Z_t^{(X;I)}\right) \in R,$$

while in (2.22), $S_t^{(MW)}, t \in [0, T]$, is the stock price with a potential bankruptcy event $\tau \in [0, T]$.

## 3. The resolution of the puzzle in classical binomial option pricing

Before considering binomial pricing in Bachelier's market model, we will discuss current developments related to the binomial pricing model in the Black-Scholes-Merton market model and the resolution of the binomial option price puzzle.

The classical binomial pricing model [34–36] sets up the binomial option price puzzle by stating that the option price is independent of the instantaneous mean returns, denoted as $\mu$. However, this leads to a conundrum: regardless of how high $\mu$ is, the binomial call option's price remains the same. In continuous time, the hedger can trade continuously in time and adjust her hedge when $\mu \uparrow \infty$, which indeed is not possible in real trading, but this does not explain why the call option's value, with a given strike, remains the same regardless of how large $\mu$ is in binomial option pricing, where the trades occur in discrete time.

Hu et al. [12–14] resolved this binomial option price puzzle by applying various extensions of the binomial option pricing models. In this section, we will briefly summarize the approach taken by Hu et al. [12–14] in the simplest setting, which motivates applying the same methodology to binomial option pricing in Bachelier's market model.

In the simplest binomial option pricing model [14, 37, 38], the risky asset (stock) $\mathcal{S}$ follows a binomial pricing tree.

Next, we make the additional assumptions that all terms in $u_\Delta$ and $d_\Delta$ of order $o\left(\frac{1}{n}\right)$ as $n \uparrow \infty$ are zero[10]:

$$(3.1) \qquad S_{(k+1)\Delta,n} = \begin{cases} S^{(u)}_{(k+1)\Delta,n} = S_{k\Delta,n}(1 + U_\Delta), \ if \ \xi_{k+1,n} = 1, \\ S^{(d)}_{(k+1)\Delta,n} = S_{k\Delta,n}(1 + D_\Delta), \ if \ \xi_{k+1,n} = 0, \end{cases}$$

where $(i)$ $S_{k\Delta,n}, k = 0,1, \dots, n, n \in \mathcal{N} = \{1,2, \dots \}$, is the stock price at time $k\Delta, k = 0,1, \dots, n, \Delta = \Delta_n = \frac{T}{n}$, $T$ is the fixed terminal time, and $S_0 > 0$. $(ii)$ For every $n \in \mathcal{N}, \xi_{k,n}, k = 1,2, \dots, n$, represents independent identically distributed (iid) Bernoulli random variables with $\mathbb{P}(\xi_{k,n} = 1) = 1 - \mathbb{P}(\xi_{k,n} = 0) = p_n$, determining the filtration $\mathbb{F}^{(n)} = \{\mathcal{F}^{(n)}_k = \sigma(\xi_{j,n}, j = 1, \dots, k), k = 1, \dots, n, \mathcal{F}^{(n)}_0 = \{\emptyset, \Omega\}, \xi_{0,n} = 0\}$ and the stochastic basis $\{\Omega, \mathbb{F}^{(n)}, \mathbb{P}\}$ on the complete probability space $(\Omega, \mathcal{F}, \mathbb{P})$. $(iii)$ $r \geq 0$ is the instantaneous riskless rate of the riskless bank account $\mathcal{B}$, with price dynamics given by

$$(3.2) \qquad b_{(k+1)\Delta,n} = b_{k\Delta,n}(1 + r\Delta), k = 0,1, \dots, n-1, b_{0,n} = 1.$$

The stock arithmetic returns are defined as follows[11]:

---

[10] This condition is necessary when transitioning from binomial option pricing to continuous-time option pricing by applying the Donsker-Prokhorov invariance principle. Please also refer to the next section, Section 4, for further details.

[11] The debate over whether to use log-returns or arithmetic (simple) returns is still ongoing, as discussed by Dorfleitner [39], Meucci [40], Hudson and Gregoriou [41], and Jamison [42]. It is worth noting that arithmetic returns are commonly used in (i) one-period investment analysis [43, p. 39], (ii) the CAPM [44], (iii) the Sharpe ratio [45], (iv) short-term riskless processes in continuous time [17, p. 106], and (iv) portfolios and consumption in continuous time, according to Duffie [17, p. 208]. On the other hand, log-returns are typically used in continuous-time option pricing, as highlighted by Shreve [38]. It is important to note that in binomial option pricing, the appropriate approach is to use arithmetic (simple) returns, as stated by Shreve [37] and Hu et al. [10]; this is also the approach taken in this section. Miskolczi [46] reached the following conclusion: "It is important to use the same type of return within one study, and one must be aware of the possible instabilities when comparing return results."

(3.3) $\qquad r_{(k+1)\Delta,n} = \frac{S_{(k+1)\Delta,n} - S_{k\Delta,n}}{S_{k\Delta,n}}, k = 0, \ldots, n-1, r_{0,n} = 0.$

Then,

$$(3.4) \qquad r_{(k+1)\Delta,n} = \begin{cases} r_{(k+1)\Delta,n}^{(u)} = u_\Delta, & if \; \xi_{k+1,n} = 1, \\ r_{(k+1)\Delta,n}^{(d)} = d_\Delta, & if \; \xi_{k+1,n} = 0. \end{cases}$$

Following the exposition in [12–14], suppose that the instantaneous mean $\mu$ and instantaneous variance $\sigma^2$ of the returns are known[12], that is,

$$(3.5) \qquad \mathbb{E}(r_{(k+1)\Delta,n}) = \mu\Delta, var(r_{(k+1)\Delta,n}) = \sigma^2\Delta.$$

From (3.4) and (3.5), it follows that

$$(3.6) \qquad u_\Delta = \mu\Delta + \sqrt{\frac{1-p_n}{p_n}}\sigma\sqrt{\Delta}, d_\Delta = \mu\Delta - \sqrt{\frac{p_n}{1-p_n}}\sigma\sqrt{\Delta}.$$

Consider a European contingent claim (ECC) (also called an option) $\mathcal{C}$, with a terminal value $C_T = g(S_T)$ and price dynamics $C_{k\Delta,n} = C(S_{k\Delta,n}, k\Delta), k = 0, \ldots, n-1.$ Then, $C_{k\Delta,n}$ is determined by the risk-neutral pricing tree:

$$(3.7) \qquad C_{k\Delta,n} = \frac{1}{1+r\Delta}\Big(q_n C_{(k+1)\Delta,n}^{(u)} + (1-q_n)C_{(k+1)\Delta,n}^{(d)}\Big),$$

with

$$(3.8) \qquad C_{(k+1)\Delta,n} = \begin{cases} C_{(k+1)\Delta,n}^{(u)} & if \; \eta_{k+1,n} = 1, \\ C_{(k+1)\Delta,n}^{(d)}, & if \; \eta_{k+1,n} = 0. \end{cases}$$

In (3.8), $\eta_{k,n}, k = 1,2,\ldots,n$, represents iid Bernoulli random variables with risk-neutral probabilities $\mathbb{P}(\eta_{k,n} = 1) = 1 - \mathbb{P}(\eta_{k,n} = 0) = q_n$, where

$$(3.9) \qquad q_n = p_n - \theta\sqrt{p_n(1-p_n)\Delta},$$

and $\theta = \frac{\mu-r}{\sigma}$ is the market price of risk. The no-arbitrage condition is $q_n \in (0,1)$ for all $n \in \mathcal{N}$. As the risk-neutral probabilities are dependent on $\mu$, the price of the call option is also dependent on $\mu$. Furthermore, in [34–36] the binomial option price is independent of $p_n$ and thus, regardless of how close $p_n$ is to 1 or 0, the call option price will be the same. According to (3.9), the risk-neutral probability $q_n$ is dependent on $p_n$, and so the option price will also be dependent on $p_n$, as shown in [12–14].

---

[12] In real applications, $\mu$ and $\sigma^2$ are estimated from historical asset return data, as in [12–14].

As $n \uparrow \infty$ and $\Delta = \Delta_n = \frac{T}{n} \downarrow 0$, the pricing tree represented by (3.1) generates a discrete price process with values in the Skorokhod space $\mathfrak{D}[0, T]$, which converges weakly to a geometric Brownian motion

$$(3.10) \qquad S_t = S_0 e^{\left(\mu - \frac{\sigma^2}{2}\right)t + \sigma B_t}, t \in [0, T],$$

where $B_t, t \in [0, T]$, is the Brownian motion on $(\Omega, \mathcal{F}, \mathbb{F}, \mathbb{P})$. In the risk-neutral world, replacing $p_n$ with $q_n$ in (3.1) causes the risk-neutral price process to converge weakly in $\mathfrak{D}[0, T]$ to

$$(3.11) \qquad S_t = S_0 e^{\left(r - \frac{\sigma^2}{2}\right)t + \sigma B_t^{\mathbb{Q}}}, t \in [0, T],$$

where $B_t^{\mathbb{Q}}, t \in [0, T]$, is the Brownian motion on $(\Omega, \mathcal{F}, \mathbb{F}, \mathbb{Q})$. In continuous time, all information about $p_n$ and $\mu$ is lost due to the presumed ability of the hedger to trade continuously over time. Disallowing continuous-time trading enables the creation of no-arbitrage financial markets without the use of semimartingales [47], thereby extending the fundamental asset pricing theorem [48, 49]. On the other hand, extending binomial option pricing to binary option pricing enables no-arbitrage option pricing processes when the underlying return process is $AR(n)$ or $MA(n), n \in \mathcal{N}$ [50, 51]. In particular, a binary tree option pricing model with a return process following $MA(1)$ embeds Roll's market microstructure model [52, 53] within the dynamic asset pricing theory [54].

## 4. Binomial option pricing in Bachelier's market model

There are two reasons that binomial option pricing is an important part of Bachelier's market model. The first is that discrete-time option pricing will preserve the instantaneous mean return and the upward and downward probabilities of the pricing tree in the option price, similar to the results in the previous section. Second, taking the limit to obtain the continuous-time Bachelier option pricing should justify the price dynamics in $\left(\mathcal{A}, \mathfrak{B}^{(SIA)}, \mathcal{C}\right)$, as defined in (2.9), (2.10), and (2.12).

We start with the simplest form of binomial option pricing in Bachelier's discrete-time market model $\left(\mathcal{A}^n, \mathfrak{B}^{n(SIA)}, \mathcal{C}^n\right), n \in \mathcal{N}$. The risky asset $\mathcal{A}^n$ follows Bachelier's binomial pricing tree, which has the following form:

$$(4.1) \qquad A_{(k+1)\Delta,n} = \begin{cases} A_{(k+1)\Delta,n}^{(u)} = A_{k\Delta,n} + u_\Delta, \ if \ \xi_{k+1,n} = 1, \\ A_{(k+1)\Delta,n}^{(d)} = A_{k\Delta,n} + d_\Delta, \ if \ \xi_{k+1,n} = 0, \end{cases}$$

where $(i)$ $A_{k\Delta,n}, k = 0,1, \dots, n, n \in \mathcal{N} = \{1,2, \dots\}$, is Bachelier's stock price at time $k\Delta, k = 0,1, \dots, n, \Delta = \Delta_n = \frac{T}{n}, T$ is the fixed terminal time, and $A_0 > 0$. $(ii)$ For every $n \in \mathcal{N}, \xi_{k,n}, k = 1,2, \dots, n$, represents iid Bernoulli random variables with $\mathbb{P}(\xi_{k,n} = 1) = 1 - \mathbb{P}(\xi_{k,n} = 0) = p_n$, determining the filtration $\mathbb{F}^{(n)} = \left\{ \mathcal{F}_k^{(n)} = \sigma(\xi_{j,n}, j = 1, \dots, k), k = 1, \dots, n, \mathcal{F}_0^{(n)} = \{\emptyset, \Omega\}, \xi_{0,n} = 0 \right\}$ and the stochastic basis $\left\{\Omega, \mathcal{F}, \mathbb{F}^{(n)}, \mathbb{P}\right\}$ on the complete probability space $(\Omega, \mathcal{F}, \mathbb{P})$. $(iii)$ $r^{(si)}$ is the instantaneous simple riskless rate of the simple riskless account $\mathfrak{B}^{n(SIA)}$, with price dynamics given by

$$(4.2) \qquad \beta_{(k+1)\Delta,n} = \beta_{k\Delta,n} + r^{(si)}\Delta = \beta_{0,n} + (k+1)r^{(si)}\Delta, k = 0,1, \dots, n-1, \beta_{0,n} > 0.$$

Bachelier's stock price changes are defined as follows:

$$(4.3) \qquad c_{(k+1)\Delta,n} = A_{(k+1)\Delta,n} - A_{k\Delta,n}, k = 0, \dots, n-1, c_{0,n} = 0.$$

Then,

$$(4.4) \qquad c_{(k+1)\Delta,n} = \begin{cases} c_{(k+1)\Delta,n}^{(u)} = u_\Delta, \ if \ \xi_{k+1,n} = 1, \\ c_{(k+1)\Delta,n}^{(d)} = d_\Delta, \ if \ \xi_{k+1,n} = 0. \end{cases}$$

Suppose that the instantaneous mean $\rho$ and instantaneous variance $v^2$ of Bachelier's stock price changes are known, that is,

$$(4.5) \qquad \mathbb{E}(c_{(k+1)\Delta,n}) = \rho\Delta, var(c_{(k+1)\Delta,n}) = v^2\Delta.$$

Next, we make the additional assumptions that all terms in $u_\Delta$ and $d_\Delta$ of order $o\left(\frac{1}{n}\right)$ as $n \uparrow \infty$ are zero, as required by the Donsker-Prokhorov invariance principle (DPIP), when taking the limit as $n \uparrow \infty$ and $\Delta = \Delta_n = \frac{T}{n} \downarrow 0$. Then, $u_\Delta$ and $d_\Delta$ are determined by

(4.6)     $u_\Delta p_n + d_\Delta (1 - p_n) = \rho\Delta, \ u_\Delta^2 p_n + d_\Delta^2 (1 - p_n) = v^2 \Delta.$

According to (4.6), the constants $u_\Delta$ and $d_\Delta$ are determined by $\rho, p_n,$ and $v$:

(4.7)     $u_\Delta = \rho\Delta + \sqrt{\frac{1-p_n}{p_n}} \, v\sqrt{\Delta}, \ d_\Delta = \rho\Delta - \sqrt{\frac{p_n}{1-p_n}} \, v\sqrt{\Delta}.$

As in [12, 13], the pricing tree represented by (4.1) and (4.7) generates a càdlàg price process

(4.8)   $A_t^{(n)} = A_{k\Delta,n}, if \ t \in [k\Delta, (k+1)), k = 0,1,\dots, n-1, A_T^{(n)} = A_{n\Delta}^{(n)} = A_{n\Delta,n} = A_T,$

in the Skorokhod space $\mathfrak{D}[0,T]$. As $n \uparrow \infty$, the process $A_t^{(n)}, t \in [0,T]$, converges weakly to the arithmetic Brownian motion $A_t, t \in [0,T]$, given by (2.9), that is,

(4.9)         $dA_t = \rho dt + v dB_t, t \in [0,T], A_0 > 0, \rho \in R, v > 0.$

The SIA in (4.2) of $\mathfrak{B}^{n(SIA)}$ generates a deterministic càdlàg process

(4.10)   $\beta_t^{(n)} = \beta_{k\Delta,n}, if \ t \in [k\Delta, (k+1)), k = 0,1,\dots, n-1, \beta_T^{(n)} = \beta_{n\Delta}^{(n)} = \beta_{n\Delta,n} = \beta_T.$

As $n \uparrow \infty$, the process $\beta_t^{(n)}, t \in [0,T]$, converges in the Skoroknod topology in $\mathfrak{D}[0,T]$ to $\beta_t, t \in [0,T]$, which is given by (2.10), that is,

(4.11)         $d\beta_t = r^{(si)} dt, t \in [0,T], \beta_0 > 0, r^{(si)} \in (-\infty, \rho).$

Consider an option (ECC) $\mathcal{G}^{(n)}$, with a terminal value $G_T^{(n)} = g\left(\beta_T^{(n)}\right)$ and price dynamics $G_{k\Delta,n} = D\left(A_{k\Delta,n}, k\Delta\right), k = 0,\dots, n-1.$ Suppose the hedger taking a short position in $\mathcal{G}^{(n)}$ chooses to form a riskless Bachelier's portfolio $P_{k\Delta,n} = a_{k\Delta,n} A_{k\Delta,n} - G_{k\Delta,n}.$ Given $\mathcal{F}_k^{(n)},$ $P_{(k+1)\Delta,n} = a_{k\Delta,n} A_{(k+1)\Delta,n} - G_{(k+1)\Delta,n},$ that is,

$P_{(k+1)\Delta,n}^{(u)} = a_{k\Delta,n} A_{(k+1)\Delta,n}^{(u)} - G_{(k+1)\Delta,n}^{(u)} = P_{(k+1)\Delta,n}^{(d)} = a_{k\Delta,n} A_{(k+1)\Delta,n}^{(d)} - G_{(k+1)\Delta,n}^{(d)}.$

Thus,

$$(4.12) \qquad a_{k\Delta,n} = \frac{G^{(u)}_{(k+1)\Delta,n} - G^{(d)}_{(k+1)\Delta,n}}{A^{(u)}_{(k+1)\Delta,n} - A^{(d)}_{(k+1)\Delta,n}} = \frac{G^{(u)}_{(k+1)\Delta,n} - G^{(d)}_{(k+1)\Delta,n}}{u_\Delta - d_\Delta}.$$

The hedger's Bachelier's portfolio is riskless, which implies that

$$(4.13) \quad a_{k\Delta,n} A_{k\Delta,n} - G_{k\Delta,n} = P_{k\Delta,n} = P_{(k+1)\Delta,n} - r^{(si)}\Delta = a_{k\Delta,n} A^{(u)}_{(k+1)\Delta,n} - G^{(u)}_{(k+1)\Delta,n} - r^{(si)}\Delta.$$

Now, (4.7), (4.12), and (4.13) lead to Bachelier's binomial option pricing

$$(4.14) \qquad G_{k\Delta,n} = q^{(b)}_n G^{(u)}_{(k+1)\Delta,n} + \left(1 - q^{(b)}_n\right) G^{(d)}_{(k+1)\Delta,n} + r^{(si)}\Delta,$$

where

$$(4.15) \qquad q^{(b)}_n = p_n - \frac{\rho}{v}\sqrt{p_n(1-p_n)}\sqrt{\Delta}\,,\, 1 - q^{(b)}_n = 1 - p_n + \frac{\rho}{v}\sqrt{p_n(1-p_n)}\sqrt{\Delta},$$

represents Bachelier's risk-neutral probabilities. The no-arbitrage condition for Bachelier's market model $\left(\mathcal{A}^n, \mathfrak{B}^{n(SIA)}, \mathcal{C}^n\right), n \in \mathcal{N}$, is $q^{(b)}_n \in (0,1)$ for all $n \in \mathcal{N}$.

Next, consider Bachelier's stock price changes

$$c^{(\mathbb{Q})}_{(k+1)\Delta,n} = A^{(\mathbb{Q})}_{(k+1)\Delta,n} - A^{(\mathbb{Q})}_{k\Delta,n}, k = 0, \dots, n-1, c^{(\mathbb{Q})}_{0,n} = 0, \quad A^{(\mathbb{Q})}_0 = A_0 > 0,$$

in the risk-neutral world $\mathbb{Q}$, determined by the risk-neutral probabilities represented by (4.15). Then,

$$(4.16) \qquad c^{(\mathbb{Q})}_{(k+1)\Delta,n} = \begin{cases} c^{(\mathbb{Q},u)}_{(k+1)\Delta,n} = u_\Delta, \; if \; \eta_{k+1,n} = 1, \\ c^{(\mathbb{Q},d)}_{(k+1)\Delta,n} = d_\Delta, \; if \; \eta_{k+1,n} = 0, \end{cases}$$

where $\mathbb{P}\left(\eta_{k+1,n} = 1\right) = q^{(b)}_n$. From (4.14) and the assumption $o(\Delta) = o\left(\frac{T}{n}\right) = 0$, it follows that

$$(4.17) \qquad \begin{bmatrix} \mathbb{E}^{\mathbb{Q}}\left(c_{(k+1)\Delta,n}\right) = u_\Delta q_n + d_\Delta(1-q_n) + r^{(si)}\Delta = r^{(si)}\Delta, \\ var^{\mathbb{Q}}\left(c_{(k+1)\Delta,n}\right) = var^{\mathbb{Q}}\left(c_{(k+1)\Delta,n} - r^{(si)}\Delta\right) = \sigma^2\Delta. \end{bmatrix}$$

Thus, as in (4.8) and (4.9), we have that the risk-neutral pricing tree represented by (4.14) and (4.15) generates a càdlàg price process on $\mathbb{Q}$:

$$(4.18) \quad A^{(\mathbb{Q},n)}_t = A^{(\mathbb{Q})}_{k\Delta,n}, if \; t \in \left[k\Delta, (k+1)\right), k = 0,1,\dots,n-1, A^{(\mathbb{Q},n)}_T = A^{(\mathbb{Q},n)}_{n\Delta} = A^{(\mathbb{Q})}_{n\Delta,n} = A^{(\mathbb{Q})}_T.$$

As $n \uparrow \infty$, the process $A_t^{(\mathbb{Q},n)}, t \in [0,T]$, converges weakly in the Skorokhod space $\mathfrak{D}[0,T]$ to the arithmetic Brownian motion $A_t^{(\mathbb{Q})}, t \in [0,T]$, given by

$$(4.19) \qquad dA_t^{(\mathbb{Q})} = r^{(si)}\,dt + vdB_t^{(r^{(si)})}, t \in [0,T], A_0 > 0, \rho \in R, v > 0.$$

Thus, we extended the binomial option pricing from Section 3 to option pricing in Bachelier's market model in continuous time. As in (3.7)–(3.11), Bachelier's option pricing model, represented by (4.14) and (4.15), preserves (i) the natural-world (historical) probabilities $p_n$ of the asset price's upturn in the $n$-period Bachelier's pricing tree and (ii) the instantaneous mean price change $\rho$. In the limit, as expected, that information is lost, as shown in (4.19).

## 5. Bachelier pricing model for path-dependent price changes

In this section, we follow the exposition in [12] to extend the results in the previous section to Bachelier's option pricing on a pricing tree with dependent price changes and time-dependent probabilities for upward and downward price movements.

We start with the description of Bachelier's market model $\left(\mathcal{A}^n, \mathfrak{B}^{n(SIA)}, \mathcal{C}^n\right), n \in \mathcal{N}$, with time-dependent probabilities for upward and downward price movements. On a complete probability space $(\Omega, \mathcal{F}, \mathbb{P})$, consider a triangular array of binary random variables $\varsigma_{n,1}, \varsigma_{n,2}, \ldots, \varsigma_{n,k_n}, \ n \in N$, satisfying the following conditions: (a) for every $n \in N, \varsigma_{n,1}, \varsigma_{n,2}, \ldots, \varsigma_{n,k_n}, \ n \in N$, is a sequence of independent binary random variables with $\mathbb{P}(\varsigma_{n,k} = 1) = 1 - \mathbb{P}(\varsigma_{n,k} = -1) = p_{n,k} \in (0,1), k = 1,\ldots,k_n, n \in N$. On $(\Omega, \mathcal{F}, \mathbb{P})$, consider the discrete filtration $\mathbb{F}^{(n)} = \left\{\mathcal{F}^{(n,0)} = \{\emptyset, \Omega\}, \mathcal{F}^{(n,k)} = \sigma\left(\varsigma_{n,1}, \ldots, \varsigma_{n,k}\right), k = 1,\ldots,k_n\right\}, n \in \mathcal{N}$. Let $0 = t_{n,0} < t_{n,1} < \cdots < t_{n,k} < \cdots < t_{n,k_n} = T$ be the times that the option hedger can take a short position in the option contract $\mathcal{C}^n$. The risky asset $\mathcal{A}^n$ has price dynamics determined by Bachelier's pricing tree: $A_{t_{n,0}}^{(n)} = A_0 > 0$, and for $k = 1, \ldots, k_n$,

$$(5.1) \quad A_{t_{n,k}}^{(n)} = \begin{cases} A_{t_{n,k}}^{(n,u)} = A_{t_{n,k-1}}^{(n)} + \rho\Delta t_{n,k} + v\sqrt{\dfrac{1-p_{n,k}}{p_{n,k}}}\sqrt{\Delta t_{n,k}}, \ if \ \varsigma_{n,k} = 1, \\[3mm] A_{t_{n,k}}^{(n,d)} = A_{t_{n,k-1}}^{(n)} + \rho\Delta t_{n,k} - v\sqrt{\dfrac{p_{n,k}}{1-p_{n,k}}}\sqrt{\Delta t_{n,k}}, \ if \ \varsigma_{n,k} = -1, \end{cases}$$

where $\Delta t_{n,k} = t_{n,k} - t_{n,k-1}, k = 1, \ldots, k_n$. We assume that $\max_{k=1,\ldots,k_n} \Delta t_{n,k} = O\left(\frac{1}{n}\right)$. Consider then the càdlàg price process

$$(5.2) \quad \mathbb{A}_{[0,T]}^{(n)} = \left\{ A_t^{(n)} = A_{t_{n,k}}^{(n)}, \; t \in [t_{n,k}, t_{n,k+1}), k = 0,1,\ldots,k_n-1, A_T^{(n)} = A_{t_{n,k_n}}^{(n)} = A_T \right\}.$$

By the Donsker-Prokhorov invariance principle (DPIP)[13], $\mathbb{A}_{[0,T]}^{(n)}$ converges weakly to Bachelier's price process:

$$(5.3) \quad \mathbb{A}_{[0,T]} = \{ A_t = A_0 + \rho t + v B_t, \; t \in [0,T] \},$$

where $B_t, t \in [0,T]$, is a Brownian motion generating the stochastic basis $(\Omega, \mathcal{F}, \mathbb{F} = \{ \mathcal{F}_t = \sigma(B_u, u \le t) \subseteq \mathcal{F}, t \in [0,T] \}, \mathbb{P})$ on $(\Omega, \mathcal{F}, \mathbb{P})$.

Next, let $r^{(si)} < \rho$ be the instantaneous simple riskless rate of the simple interest account (SIA) $\mathfrak{B}^{n(SIA)}$ with price dynamics given by

$$(5.4) \quad \beta_{t_{n,k}}^{(n)} = \beta_{t_{n,k-1}}^{(n)} + r^{(si)} \Delta t_{n,k}, k = 1, \ldots, n, \beta_{0,n} > 0.$$

The simple interest account (5.4) of $\mathfrak{B}^{n(SIA)}$ generates a deterministic càdlàg process

$$(5.5) \quad \beta_t^{(n)} = \beta_{t_{n,k}}^{(n)}, if \; t \in [t_{n,k}, t_{n,k+1}), \; k = 0,1,\ldots,k_n-1, \beta_T^{(n)} = \beta_{t_{n,k_n}}^{(n)} = \beta_T.$$

As $n \uparrow \infty$, the process $\beta_t^{(n)}, t \in [0,T]$, converges in the Skorokhod topology in $\mathfrak{D}[0,T]$ to $\beta_t, t \in [0,T]$, which is given by (2.10).

Using the same no-arbitrage arguments utilized in Section 4, we find that Bachelier's risk-neutral probabilities, corresponding to the tree represented by (5.1), are given by

$$(5.6) \quad q_{n,k}^{(b)} = p_{n,k} - \frac{\rho}{v}\sqrt{p_{n,k}(1-p_{n,k})\Delta t_{n,k}}, 1 - q_{n,k}^{(b)} = 1 - p_{n,k} + \frac{\rho}{v}\sqrt{p_{n,k}(1-p_{n,k})\Delta t_{n,k}}.$$

The price $G_{t_{n,k}}^{(n)}, k = 0,1,\ldots,k_n-1$, of the option $\mathcal{C}^n$ is given by

$$G_{t_{n,k}}^{(n)} = q_{n,k}^{(b)} G_{t_{n,k+1}}^{(n,u)} + \left(1 - q_{n,k}^{(b)}\right) G_{t_{n,k+1}}^{(n,d)} + r^{(si)} \Delta t_{n,k+1}.$$

---

[13] See [55–57].

Bachelier's risk-neutral pricing tree is given by

$$(5.7) \quad A_{t_{n,k}}^{(n)} = \begin{cases} A_{t_{n,k}}^{(n,u)} = A_{t_{n,k-1}}^{(n)} + \rho \Delta t_{n,k} + v \sqrt{\frac{1-p_{n,k}}{p_{n,k}}} \sqrt{\Delta t_{n,k}}, \ if \ \varsigma_{n,k}^{(q)} = 1, \\ A_{t_{n,k}}^{(n,u)} = A_{t_{n,k-1}}^{(n)} + \rho \Delta t_{n,k} - v \sqrt{\frac{p_{n,k}}{1-p_{n,k}}} \sqrt{\Delta t_{n,k}}, \ if \ \varsigma_{n,k}^{(q)} = -1, \end{cases}$$

where $\varsigma_{n,1}^{(q)}, \varsigma_{n,2}^{(q)}, \dots, \varsigma_{n,k_n}^{(q)}$, $n \in N$, is a sequence of independent binary random variables with $\mathbb{P}\left(\varsigma_{n,k}^{(q)} = 1\right) = 1 - \mathbb{P}\left(\varsigma_{n,k}^{(q)} = -1\right) = q_{n,k} \in (0,1), k = 1, \dots, k_n, n \in \mathcal{N}$.

We now consider the extension of Bachelier's binomial option pricing ((5.1)–(5.7)) to the case in which the price changes

$$(5.8) \qquad c_{t_{n,k}}^{(n)} = A_{t_{n,k}}^{(n)} - A_{t_{n,k-1}}^{(n)}, k = 1, \dots, k_n, c_{t_{n,0}}^{(n)} = 0,$$

are dependent. We will introduce a Bachelier's pricing tree in which the random signs, representing the sign of the price changes $c_{t_{n,k}}^{(n)}, k = 1, \dots, k_n$, are determined by the price changes of some risky factor $\mathfrak{F}^n$ affecting the price of Bachelier's asset $\mathcal{A}^{n}$[14]. To take the weak limits as in (5.1)–(5.3) and obtain a continuous-time Bachelier's pricing model with time-dependent price changes, we will need the Cherny-Shiryaev-Yor [59][15] extension of the DPIP, which we will formulate now.

Let $\xi_k, k \in \mathcal{N}$, be a sequence of iid random variables with mean 0 and variance 1. Set $\xi_0^{(n)} = 0, \xi_k^{(n)} = n^{-\frac{1}{2}}\xi_k$, and $X_{\frac{k}{n}}^{(n)} = \sum_{i=1}^{k} \xi_i^{(n)}, k \in \mathcal{N}, n \in \mathcal{N}$. Let $\mathbb{B}_t^{(n)}, t \geq 0$, be the random process

---

with piecewise linear trajectories with vertices $\left(\frac{k}{n}, \mathbb{B}_{\frac{k}{n}}^{(n)}\right)$, where $\mathbb{B}_{\frac{k}{n}}^{(n)} = X_{\frac{k}{n}}^{(n)}, k \in \mathcal{N}_0 = \{0, 1, 2, \dots\}, n \in \mathcal{N}$. Following Cherny et al. [59], call a function $h: R \to R$ a *CSY-piecewise continuous function* if there exists a collection of disjoint intervals $J_n, n \in \mathcal{N}$[16], such that (a) $\bigcup_{n=1}^{\infty} J_n = R$, (b) every compact interval on $R$ will be covered by finite numbers of $J_n, n \in \mathcal{N}$, and (c) $h: J_n \to R$ is continuous and has finite limits at those endpoints of $J_n$ that do not belong to $J_n$. Let $Y_{\frac{k}{n}}^{(n)} = \sum_{i=1}^{k} h\left(X_{\frac{i-1}{n}}^{(n)}\right)\xi_i^{(n)}, k \in \mathcal{N}, n \in \mathcal{N}$. For every fixed $n \in \mathcal{N}$, define $\mathbb{C}_t^{(n)}, t \geq 0$, to be the random process with CSY-piecewise linear trajectories with vertices $\left(\frac{k}{n}, \mathbb{C}_{\frac{k}{n}}^{(n)}\right)$, where $\mathbb{C}_{\frac{k}{n}}^{(n)} = Y_{\frac{k}{n}}^{(n)}, k \in \mathcal{N}, n \in \mathcal{N}$.

**Cherny-Shiryaev-Yor invariance principle (CSYIP)**[17]: *Let $h: R \to R$ be a CSY-piecewise continuous function. Then, as $n \uparrow \infty$, the bivariate process $\left(\mathbb{B}_t^{(n)}, \mathbb{C}_t^{(n)}\right), t \geq 0$, converges in law[18] to $(B_t, C_t), t \geq 0$, where $B_t, t \geq 0$, is a standard Brownian motion and $C_t = \int_0^t h(B_s)ds, t \geq 0$.*

Examples of applications of the CSYIP include extensions of the DPIP for local time processes and skew Brownian motion, among others (see [10, 12–14, 59]). In Appendix B, we introduce a generalized skew Brownian motion based on the CSYIP.

Following (5.1)–(5.8), we now consider the dynamics of the historical (observed) factor price process $A_{t_{n,k}}^{(n,F)}, t_{n,k} = k\Delta, \ \Delta = \Delta_n = \frac{T}{n}, n \in \mathcal{N}, A_{t_{n,0}}^{(n,F)} = A_0^{(F)} > 0$. Let

(5.9) $\qquad c_{t_{n,k}}^{(n,F,hist)} = A_{t_{n,k}}^{(n,F,hist)} - A_{t_{n,k-1}}^{(n,F,hist)}, k = 1, \dots, n, c_{t_{n,0}}^{(n,F,hist)} = 0,$

---

[16] Each $J_n$ can be a closed, open, or semi-open interval or a point.

[17] In [12], an extension of the CSYIP is used.

[18] Convergence in law corresponds to the weak convergence of probability measures on $C([0, \infty), R)$ with the topology of the uniform convergence on compact intervals; see also subsection 2.2, page 45 of [60].

be the observed (historical) factor price changes. Let $\varsigma_{n,k}^{(F)}$ be a sign random variable: $\varsigma_{n,k}^{(F)} = 1$ if $c_{t_{n,k}}^{(n,F)} \geq 0$, and $\varsigma_{n,k}^{(F)} = -1$ if $c_{t_{n,k}}^{(n,F)} < 0$. We assume that $\varsigma_{n,k}^{(F)}, k = 1, \ldots, n$, represents independent random variables, with $\mathbb{P}\left(\varsigma_{n,k}^{(F)} = 1\right) = 1 - \mathbb{P}\left(\varsigma_{n,k}^{(F)} = -1\right) = p_{n,k}^{(F)} = p_0^{(F)} + p_1^{(F)}\sqrt{\Delta} + p_2^{(F)}\Delta$, for some $p_0^{(F)} \in (0,1), p_1^{(F)} \in R, p_2^{(F)} \in R^{19}$. We assume that $n \in \mathcal{N}$ is large enough that $p_{n,k}^{(F)} \in (0,1)$. The special form for $p_{n,k}^{(F)} = p_0^{(F)} + p_1^{(F)}\sqrt{\Delta} + p_2^{(F)}\Delta$ is needed when we take the limit as $n \uparrow \infty$ in order to apply the CSYIP. We assume that the dynamics of the price changes given in (5.9) are determined by Bachelier's pricing tree for the factor $\mathfrak{F}^n$: $A_{t_{n,0}}^{(n,F)} = A_0^{(F)} > 0, c_{t_{n,0}}^{(n,F)} = 0$, and for $k = 1, \ldots, n$,

$$(5.10) \qquad A_{t_{n,k}}^{(n,F)} = A_{t_{n,k-1}}^{(n,F)} + c_{t_{n,k}}^{(n,F)}, c_{t_{n,k}}^{(n,F)} = \begin{cases} \rho^{(F)}\Delta t_{n,k} + v^{(F)}\sqrt{\dfrac{1-p_{n,k}}{p_{n,k}}}\sqrt{\Delta t_{n,k}}, & if\ \varsigma_{n,k}^{(F)} = 1, \\[4mm] \rho^{(F)}\Delta t_{n,k} - v^{(F)}\sqrt{\dfrac{p_{n,k}}{1-p_{n,k}}}\sqrt{\Delta t_{n,k}}, & if\ \varsigma_{n,k}^{(F)} = -1. \end{cases}$$

The parameters $\rho^{(F)} > r^{(si)}$ and $v^{(F)} > 0$ are estimated from historical price data on $\mathfrak{F}^n$ fitted to the price dynamics described by (5.10). Consider the stochastic process with trajectories in $\mathfrak{D}[0,T]$ generated by Bachelier's price tree (5.10):

$$(5.11)\ \mathbb{A}_{[0,T]}^{(n,F)} = \left\{ A_t^{(n,F)} = A_{t_{n,k}}^{(n,F)}, \ t \in [t_{n,k}, t_{n,k+1}), k = 0,1,\ldots,k_n - 1, A_T^{(n,F)} = A_{t_{n,k_n}}^{(n,F)} = A_T^{(F)} \right\}.$$

By the DPIP, $\mathbb{A}_{[0,T]}^{(n,F)}$ converges weakly in the Skorokhod topology on $\mathfrak{D}[0,T]$ to Bachelier's arithmetic Brownian motion:

$$(5.12) \qquad \mathbb{A}_{[0,T]}^{(F)} = \left\{ A_t^{(F)} = A_0^{(F)} + \rho^{(F)}t + v^{(F)}B_t, \ t \in [0,T] \right\}.$$

Next, we would like to take the dynamics of ups and downs in the factor price process and linearly transform them to obtain the appropriate driving sequence of ups and downs for the underlying asset of the option contract. Due to the general form of the CSYIP, the underlying asset price dynamics (in discrete and continuous time) will have a flexible distribution to

---

[19] $p_0^{(F)}, p_1^{(F)}$, and $p_2^{(F)}$ should be estimated from a sample of the historical factor price changes $c_{t_{n,k}}^{(n,F,hist)}$.

accommodate some of the stylized facts of asset prices[20]. To this end, we will apply the CSYIP by following the set of transformations linking the factor price dynamics with those of the asset:

(Step 1) Strip the dynamics of the factor's ups and downs from the discrete-time factor price changes by setting

(5.13)   $Z_{t_{n,k}}^{(n,F)} = \frac{1}{v^{(F)}\sqrt{\Delta}}\left(c_{t_{n,k}}^{(n,F)} - \rho^{(F)}\Delta\right).$

By (5.10), $Z_{t_{n,k}}^{(n,F)}$ represents independent random variables with zero mean and variance 1.

(Step 2) Use the ups and downs of $Z_{t_{n,k}}^{(n,F)}$ to form the sequence of iid random signs with zero mean and variance 1, as required by the CSYIP:

(5.14) $\xi_{n,k}^{(n,F)} = \sqrt{\frac{1-p_{n,k}^{(F)}}{p_{n,k}^{(F)}}}I_{\left\{Z_{t_{n,k}}^{(n,F)}\geq 0\right\}} - \sqrt{\frac{p_{n,k}^{(F)}}{1-p_{n,k}^{(F)}}}I_{\left\{Z_{t_{n,k}}^{(n,F)}<0\right\}},$

where $p_{n,k}^{(F)} = \mathbb{P}\left(Z_{t_{n,k}}^{(n,F)} \geq 0\right)$ is the probability of an upturn in the factor's centralized price change.

(Step 3) Form the underlying discrete processes in the CSYIP, which must be properly normalized to obtain the necessary version of the CSYIP on $\mathfrak{D}[0,T]$:

(5.15) $X_{\frac{k}{n}}^{(n)} = \sum_{i=1}^{k} \xi_{n,i}^{(n,F)}\sqrt{\Delta}, Y_{\frac{k}{n}}^{(n,h)} = \sum_{i=1}^{k} \xi_{n,i}^{(n,F)} h\left(X_{\frac{i-1}{n}}^{(n)}\right)\sqrt{\Delta}.$

As with the derivation of (5.1), (5.2), and (5.3), we can embed $\left(X_{\frac{k}{n}}^{(n)}, Y_{\frac{k}{n}}^{(n,h)}\right)_{k=0,\dots,n}$ into a bivariate process with trajectories in $\mathfrak{D}[0,T] \times \mathfrak{D}[0,T]$, which will converge weakly to $\left(\mathbb{B}_{[0,T]}, \mathbb{C}_{[0,T]}\right)$, where $\mathbb{B}_{[0,T]}$ is a Brownian motion $B_t, t \in [0,T]$, and $\mathbb{C}_{[0,T]} = \Big\{C_t^{(h)} = \int_0^t h(B_s)dB_s\Big\}_{t\in[0,T]}.$

---

[20] See Rachev and Mittnik [61, Introduction] and Cont [62].

Next, similarly to (5.2), we define the price dynamics $A_t^{(n,h)}, t \in [0,T]$, of Bachelier's asset $\mathcal{A}^n$ as a $\mathfrak{D}[0,T]$-process with $A_{t_0}^{(n,h)} = A_0 > 0$ and price changes $c_{t_k}^{(n,h)}, k = 1, \ldots, n$, that depend on the past factor price changes:

$$(5.16) \qquad c_{t_k}^{(n,h)} = A_{t_k}^{(n,h)} - A_{t_{k-1}}^{(n,h)} = \rho^{(\mathcal{A})}\Delta + v^{(\mathcal{A})}\sqrt{\Delta}\xi_{n,k}^{(n,F)} + \gamma^{(\mathcal{A})}\sqrt{\Delta}\xi_{n,k}^{(n,F)}h\left(\sum_{i=1}^{k-1}\sqrt{\Delta}\xi_{n,i}^{(n,F)}\right),$$

where $\rho^{(\mathcal{A})} \in R, v^{(\mathcal{A})} \in R,$ and $\gamma^{(\mathcal{A})} \in R^{21}$ are parameters to be estimated from historical price data on $\mathcal{A}^n$. The derivation of the risk-neutral dynamics of $A_t^{(n,h)}, t \in [0,T]$, follows the same arguments utilized in Section 4 and Section 5 in [12], and thus it is omitted.

## 6. Conclusion

Over the past two decades, ESG investing and ESG finance have fundamentally reshaped how researchers and practitioners perceive the investment process, challenging the traditional risk-return investment theory and practice. In our approach to ESG investing and ESG finance, as presented in [9] and [10], we consider the ESG savviness of investors as a third dimension alongside the risk dimension and return dimension in the investment process.

In this paper, we have proposed Bachelier's market model as a natural framework for incorporating ESG-adjusted stock prices. Our approach involves introducing a simple interest rate account as the riskless asset in Bachelier's market model. By applying the Donsker-Prokhorov invariance principle and the more general Cherny-Shiryaev-Yor invariance principle, we have derived Bachelier's option pricing models in both discrete and continuous time.

One notable feature of our discrete option pricing models is that they preserve the parameters of the underlying spot prices. As a result, these parameters can be calibrated using existing market option prices. However, this calibration analysis is also an avenue for future research.

---

[21] Note that $v^{(\mathcal{A})}$ and $\gamma^{(\mathcal{A})}$ can take any real values, as the signs of the price changes for the factor $\mathfrak{F}^n$ and the asset $\mathcal{A}^n$ can take opposite values.

Overall, our study contributes to the understanding of incorporating ESG considerations into financial market models and option pricing. It opens up possibilities for further exploration and empirical validation in the field of ESG finance.

## 7. **Appendix A The ESG-adjusted stock price within Bachelier's market model**

We assumed in Section 2 that the risky asset $\mathcal{A}$ has the price dynamics of an arithmetic Brownian motion

$$A_t = A_0 + \rho t + v B_t, t \in [0, T], A_0 \in R, \rho \in R, v > 0.$$

The foremost motivation for the use of the risky asset $\mathcal{A}$ in Bachelier's market model is ESG investing[22] [63–66]. ESG investing is also known as green investing[23]. An ESG investor (a green investor) quantitatively measures the ESG effort using the ESG score of the company issuing the stock with a price $S_t, t \geq 0$[24]. Some major providers of the ESG scores of financial firms are

---

[22] From MSCI (https://www.msci.com/esg-101-what-is-esg): "At MSCI, we define ESG Investing as the consideration of environmental, social and governance factors alongside financial factors in the investment decision-making process. Remy Briand, Managing Director, MSCI ESG Research."

[23] From Investopedia (https://www.investopedia.com/terms/g/green-investing.asp): "Green investing seeks to support business practices that have a favorable impact on the natural environment. Often grouped with socially responsible investing (SRI) or environmental, social, and governance (ESG) criteria, green investments focus on companies or projects committed to the conservation of natural resources, pollution reduction, or other environmentally conscious business practices. Green investments may fit under the umbrella of SRI but are more specific."

[24] From Sean Michael Kerner (ESG strategy and management: Complete guide for businesses, Tech Accelerator, https://www.techtarget.com/sustainability/definition/ESG-score): "An ESG score is a way to assign a quantitative metric, such as a numerical score or letter rating, to the environmental, social and governance (ESG) efforts undertaken by a specific organization. ESG efforts have become increasingly important in recent years as awareness around the topic has grown. There are numerous ESG benefits for businesses that get it right, including providing competitive advantage, attracting investors, improving financial performance, building customer loyalty and helping to make a company's operations sustainable. An ESG score, also sometimes

REFINITIV[25], S&P Global[26], and Morningstar[27]. Example ESG scores of companies from the Morningstar ESG ratings are given in Table A1.

**Table A1** Best ESG companies in terms of stocks (available at https://www.investors.com/news/esg-companies-list-top-100-esg-stocks-2022/)

| Rank | Company | ESG Score |
|------|---------|-----------|
| 1 | Worthington Industries | 75.82 |
| 2 | J.B. Hunt Transport Services | 73.09 |
| 3 | Verisk Analytics | 72.79 |
| 4 | Texas Instruments[28] | 72.63 |

**October 24, 2022**

---

referred to as an ESG rating, is a way to measure how organizations are executing their ESG goals as they seek to recognize the benefits it can provide."

[25] From REFINITIV (https://www.refinitiv.com/en): Environmental, social, and governance (ESG) indices: "Our environmental, social and governance (ESG) indices help you mitigate and assess the risk of companies against ESG factors and help socially responsible investors to navigate around ESG risks."

[26] From S&P Global, Market Intelligence (https://www.spglobal.com/marketintelligence/): "Get Your Company Started on the Road to Sustainability The journey to sustainability doesn't have to be long and winding. The right analytics can help you prepare investor reporting, implement carbon reduction plans, manage climate risk, and align to Paris Agreement goals."

[27] From Morningstar Direct: ESG Data for Asset Managers (https://www.morningstar.com/): "We'll show you how you can use ESG data alongside other factors in Morningstar DirectSM or Morningstar Office.SM See what is driving a fund's Sustainability Rating using our full dashboard of sustainability metrics."

[28] The closing price on October 24, 2022, for Texas Instruments Inc. (NASADAQ:TXN) was 165.15 USD.

Nasdaq publishes ESG reports on its companies[29] and a global ESG index based on the Nasdaq Composite index (INDEXNASDAQ: IXIC).

We view the ESG scores measured in the interval [0,100] as unsatisfactory. As an example, suppose that a company $X$ has an ESG score $Z_t^{(X)} = 0$ at time t, and at time $t+1$, it improves its ESG score to $Z_{t+1}^{(X)} = 5$. The improvement is negligible. Next, suppose that a company $Y$ has an ESG score $Z_t^{(Y)} = 95$ at time $t$, and at time $t+1$, it reaches the perfect ESG score of $Z_{t+1}^{(X)} = 100$. This improvement is significant; the company must have worked hard in that time interval to reach the maximum score.

One possible way to overcome this deficiency of the ESG score scale is to use suitably chosen monotone transformations of the ESG score. One example of such a scale transformation is the exponential transformation $f^{(\exp)}(Z)$, where $Z \in [0,100]$ is the ESG score:

$$f^{(\exp)}(x; a) = \frac{e^{ax} - 1}{a}, x \in [0,100], a > 0.$$

Figure A1 shows the graph of $f^{(\exp)}(x; 0.05), x \in [0,100]$.

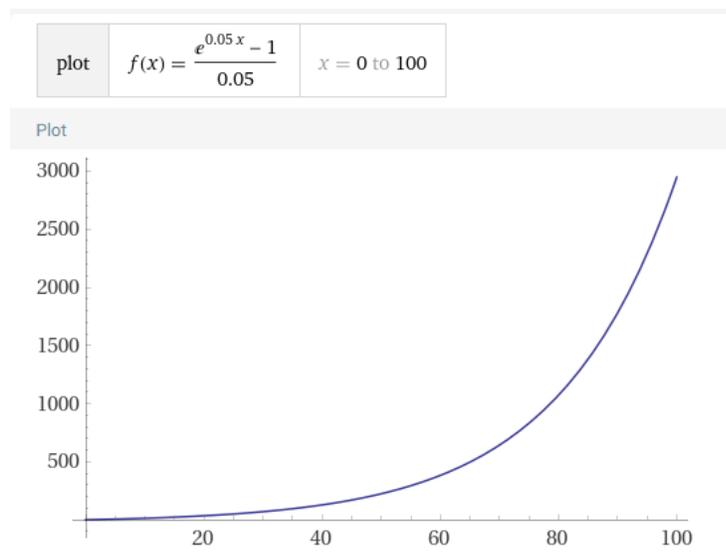

---

**Fig. A1 Exponential transformation of the standard ESG score in [0,100].**

An alternative could be the geometric function:

$$f^{(\text{geo})}(x; b) = \frac{100}{100 + b - x} - \frac{100}{100 + b}, x \in [0,100], 0 < b < 1.$$

Figure A2 shows the graph of $f^{(\text{geo})}(x; 0.5), x \in [0,100]$.

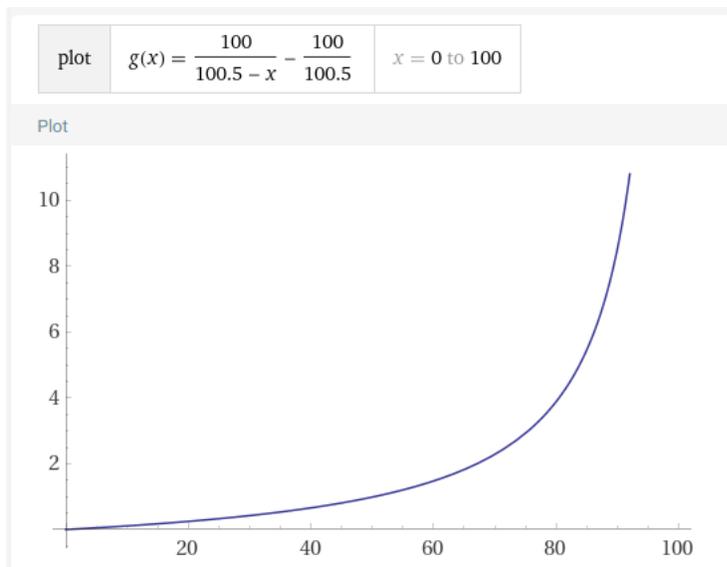

**Fig. A2 Geometric transformation of the standard ESG score in [0,100].**

Both graphs look more reasonable than the current ESG uniform scale on [0,100]. However, we believe that the choice of the monotone transformation $f(Z), Z \in [0,100]$, should be based on an axiomatic approach to ESG asset pricing, and this is beyond the scope of this paper[30].

---

[30] The transformations shown in Figures A1 and A2 reward improvements in the ESG score close to the maximum of 100 considerably more than they reward improvements close to zero. This requirement should be part of the axiomatic approach to the definition of the ESG score. This axiomatic approach to the ESG-score function should be similar to the axiomatic approach to measurable utility (see [67]).

Instead, we will use the relative ESG score with respect to the ESG index. Consider a company $X$ listed in the S&P 500 with an ESG score $Z_t^{(X)} \in [0,100]$ at time $t$[31]. Consider also the ESG S&P 500 total score $Z_t^{(I)} \in [0,100]$, which is computed using the ESG scores weighted by the stock capitalization of all companies in the S&P 500[32]. The ESG score of the S&P 500 index (INDEXSP:IND) $Z_t^{(I)}$ will serve as the ESG benchmark for $Z_t^{(X)}$. Given a fixed time $t \geq 0$, consider the **ESG performance of company $X$ relative to the ESG benchmark (also called the relative ESG score)**, which is defined as

$$\text{(A.1)} \qquad Z_t^{(X;I)} = \frac{Z_t^{(X)} - Z_t^{(I)}}{Z_t^{(I)}} \in R.$$

Suppose that the stock price of the company $X$ at time $t \geq 0$ is $S_t^{(X)} > 0$, and its relative ESG score is $Z_t^{(X;I)} \in R$. Define **the ESG-adjusted stock price of the company $X$** at time $t \geq 0$ as

$$\text{(A.2)} \qquad S_t^{ESG} = S_t^{(X)}\left(1 + \gamma^{ESG} Z_t^{(X;I)}\right) \in R.$$

In (A.2), the parameter $\gamma^{ESG} \in R$ [33] is the **ESG affinity of the financial market,** and it should be calibrated by the individual investors trading the stock of the company $X$. Green investors will place a relatively high value on $\gamma^{ESG}$ when they decide to trade the stock of the company $X$. Alternatively, $\gamma^{ESG}$ can potentially be placed in certain bounds by financial regulators.

---

[31] The ESG rating agencies typically provide the ESG rating once a year, and in some cases updates on the ESG score are provided on a monthly basis. However, there is a standard, but not perfect, procedure used to calculate the daily ESG score (see [9, 10]).

[32] From Slickcharts (https://www.slickcharts.com/sp500): "The S&P 500 component weights are listed from largest to smallest. Data for each company in the list is updated after each trading day. The S&P 500 index consists of most but not all of the largest companies in the United States. The S&P market cap is 70 to 80% of the total US stock market capitalization. It is a commonly used benchmark for stock portfolio performance in America and abroad. Beating the performance of the S&P with less risk is the goal of nearly every portfolio manager, hedge fund and private investor."

[33] While many practitioners and researchers require $\gamma^{ESG} \geq 0$ in their investment analysis, we cannot dismiss the possibility that some market participants may consider negative values for $\gamma^{ESG}$. A significant number of articles criticize the positive effect of ESG scores on optimal portfolio selection, including works by Smith [68], Bhagat [69], Webb [70], and The Economist [71]. When calibrating $\gamma^{ESG}$ from spot and derivative market prices, many involved in ESG finance, including researchers, typically assume that the worst-case scenario is investor

When Bachelier's market model is applied to the ESG financial market, the ESG-adjusted stock price $S_t^{ESG}$ is the price in (2.9):

(A.3) $S_t^{ESG} = A_t = A_0 + \rho t + vB_t, t \in [0, T], A_0 \in R, \rho \in R, v > 0.$

In Section 2, Bachelier's risky asset $\mathcal{A} = \mathcal{A}^{ESG}$ represents the ESG-adjusted stock.

## 8. Appendix B. Generalized skew Brownian motion

In this section, we will introduce the **generalized skew Brownian motion (GSBM)** based on a "horizontal-vertical" random walk [59, Section 5]. Consider the sequence of iid random variables $\xi_k, k \in \mathcal{N}$, with mean 0 and variance 1. Let us set $X_0^{(CSY)} = 0, Y_0^{(CSY)} = 0$, and for $k \in \mathcal{N}_0 = \{0, 1, \dots\}$,

$$X_{k+1}^{(CSY)} = \begin{cases} X_k^{(CSY)} + \xi_{k+1}, if \ Y_k^{(CSY)} > X_k^{(CSY)}, \\ X_k, \qquad if \ Y_k \leq X_k, \end{cases} \quad Y_{k+1} = \begin{cases} Y_k, \qquad if \ Y_k > X_k, \\ Y_k - \xi_{k+1}, if \ Y_k \leq X_k. \end{cases}$$

Consider

$$X_{\frac{k}{n}}^{(CSY;n)} = n^{-\frac{1}{2}} X_k^{(CSY)}, Y_{\frac{k}{n}}^{(CSY;n)} = n^{-\frac{1}{2}} Y_k^{(CSY)},$$

for all $k \in \mathcal{N}_0$ and $n \in \mathcal{N}$. Let $\mathbb{X}_t^{(CSY;n)}, \mathbb{Y}_t^{(CSY;n)}, \mathbb{Z}_t^{(n,\gamma,\delta)}$, be the random process with piecewise linear trajectories with vertices $\left(\frac{k}{n}, \mathbb{X}_{\frac{k}{n}}^{(CSY;n)}\right), \left(\frac{k}{n}, \mathbb{Y}_{\frac{k}{n}}^{(CSY;n)}\right)$, where $\mathbb{X}_{\frac{k}{n}}^{(CSY;n)} = X_{\frac{k}{n}}^{(CSY;n)}, \mathbb{Y}_{\frac{k}{n}}^{(CSY;n)} = Y_{\frac{k}{n}}^{(CSY;n)}, k \in \mathcal{N}_0 = \{0, 1, 2, \dots\}, n \in \mathcal{N}.$

Let $B_t, t \geq 0$, be a standard Brownian motion and let $L_t, t \geq 0$, be its local time[34]. Set

---

indifference towards ESG. However, it is possible, and perhaps even surprising, that some option traders may exhibit a general aversion to ESG which should be considered when calibrating the implied ESG affinity $\gamma^{ESG}$ at various levels of maturity and moneyness. To calculate the implied $\gamma^{ESG} \in R$ in the spot market, one should follow the standard method described by Ardia and Boudt [72].

[34] $L_t, t \geq 0$, represents the local time that the Brownian motion spends at 0 over the interval $[0, t]$, and it is defined as

$$\mathbb{X}_t^{(CSY)} = \int_0^t I_{(-\infty,0]}(B_s)\,dB_s = \frac{1}{2}L_t - \min(B_t, 0) = \frac{1}{2}\left(B_t - \int_0^t sgn(B_s)\,dB_s\right), t \geq 0,$$

$$\mathbb{Y}_t^{(CSY)} = -\int_0^t I_{(0,\infty)}(B_s)\,dB_s = \frac{1}{2}L_t - \max(B_t, 0) = \frac{1}{2}\left(-B_t - \int_0^t sgn(B_s)\,dB_s\right), t \geq 0.$$

**CSYIP for "horizontal-vertical" random walk.** As $n \uparrow \infty$, the bivariate process $\left\{\left(\mathbb{X}_t^{(CSY;n)}, \mathbb{Y}_t^{(CSY;n)}\right), t \geq 0\right\}$ converges in law to $\left\{\left(\mathbb{X}_t^{(CSY)}, \mathbb{Y}_t^{(CSY)}\right), t \geq 0\right\}$.

Consider next the following corollary of the CSYIP [59]. Let $\xi_k, k \in \mathcal{N}$, be iid random signs with $\mathbb{P}(\xi_k = 1) = \mathbb{P}(\xi_k = -1) = \frac{1}{2}$, and set

$$X_{\frac{k}{n}}^{(n)} = n^{-\frac{1}{2}}\sum_{i=1}^k \xi_i^{(n)}, L_{\frac{k}{n}}^{(n)} = n^{-\frac{1}{2}}\sum_{i=1}^k I_{\left\{\xi_i^{(n)}=0\right\}}.$$

Next, construct the processes $X_t^{(n)}, L_t^{(n)}, t \geq 0$, through the linear interpolation of $X_{\frac{k}{n}}^{(n)}, L_{\frac{k}{n}}^{(n)}, k \in \mathcal{N}_0, n \in \mathcal{N}$. Then, $\left(X_t^{(n)}, L_t^{(n)}\right), t \geq 0$, converges in law to $(B_t, L_t), t \geq 0$, as $n \uparrow \infty$. Consider $B_t^{(min)} = \frac{1}{2}L_t - \mathbb{X}_t^{(CSY)} = \min(B_t, 0)$, $B_t^{(max)} = \frac{1}{2}L_t - \mathbb{Y}_t^{(CSY)} = \max(B_t, 0)$, and consider

$$\mathbb{Z}_t^{(\gamma)} = \gamma^{(-)}B_t^{(min)} + \gamma^{(0)}L_t + \gamma^{(+)}B_t^{(max)}$$

for a given triplet of parameters $\gamma = \left(\gamma^{(-)}, \gamma^{(0)}, \gamma^{(+)}\right) \in R^3$. The graph of a trajectory of $\mathbb{Z}_t^{(1,1,1)} = B_t^{(min)} + L_t + B_t^{(max)}, 0 \leq t \leq 1$, is depicted in Figure B3.

---

$$L_t = \lim_{\varepsilon \downarrow 0}\frac{1}{2\varepsilon}\int_0^t I_{(-\varepsilon,\varepsilon)}(B_s)\,ds = \lim_{\varepsilon \downarrow 0}\frac{1}{2\varepsilon}Leb\{s: [0,t]: B_s \in (-\varepsilon,\varepsilon)\},$$

where $Leb$ is the Lebesgue measure. By Tanaka's formula,

$$|B_t| = \int_0^t sgn(B_s)\,dB_s + L_t, t \geq 0,$$

where $sgn(a)$ is $1, 0,$ or $-1$ if $a$ is greater than, equal to, or less than zero, respectively (see [73, Chapter 7]).

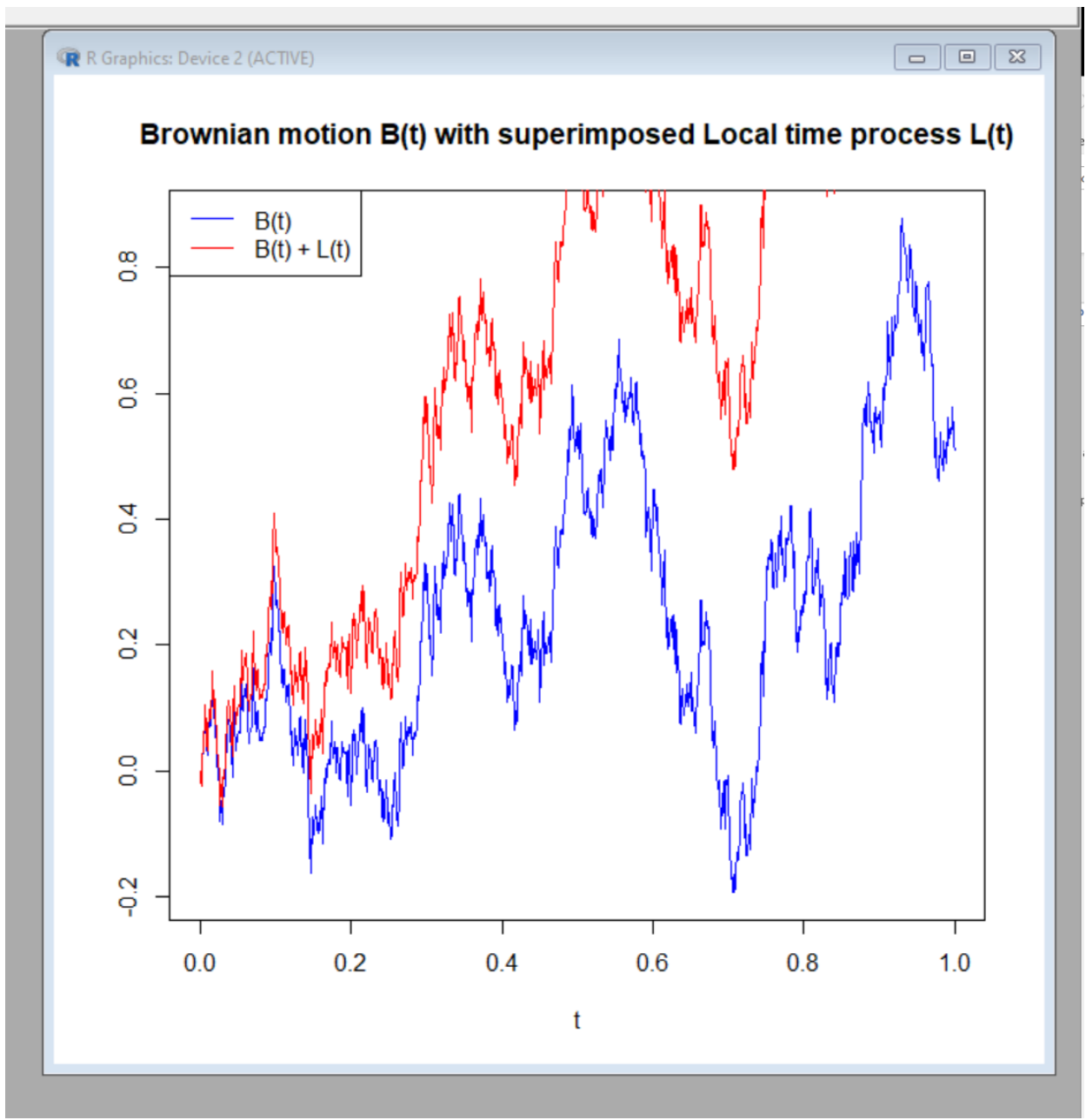

**Fig. B3 Graph of $\mathbb{Z}_t^{(1,1,1)} = B_t^{(min)} + L_t + B_t^{(max)} = B_t + L_t, t \in [0,1]$ (in red), and the Brownian motion $B_t, t \in [0,1]$ (in blue).**

Some special cases are $\mathbb{Z}_t^{((1,0,1))} = B_t, \mathbb{Z}_t^{((-1,0,1))} = |B_t|, \mathbb{Z}_t^{((0,1,0))} = L_t$, and

$$\int_0^t sgn(B_s)dB_s = |B_t| - L_t = \mathbb{Z}_t^{((-1,-1,1))}.$$

As the distribution of $\int_0^t sgn(B_s)dB_s, t \geq 0$, is that of standard Brownian motion, the skewness of the marginal distribution of $\mathbb{Z}_t^{(\gamma)}$ will be determined by $\gamma^{(-)}$ and $\gamma^{(-)}$. Black-Scholes-Merton option pricing with a price process following $dS_t = \mu S_t dt + \sigma S_t d\mathbb{Z}_t^{(\gamma)}, S_0 > 0$, and its discrete-time version can be obtained as in [74]. Similarly to Section 5, Bachelier option pricing with the asset dynamics $dA_t = \rho dt + v d\mathbb{Z}_t^{(\gamma)}, A_0 > 0$, can be readily obtained, along with its discrete time analog.

To enhance the model to exhibit the heavy tails of the marginal distribution, we consider three independent Brownian motions $B_{i,t}, i = 1,2,3$, and define the **generalized skew Brownian motion (GSBM)** as

$$\mathbb{W}_t^{(\gamma)} = \left(\gamma^{(-)} + \gamma^{(0)}\right)B_{1,t}^{(min)} - \gamma^{(0)}\int_0^t sgn(B_{2,s})dB_{2,s} + \left(\gamma^{(+)} + \gamma^{(0)}\right)B_{3,t}^{(max)}$$

$$= (in\ distribution/law) =$$

$$= \left(\gamma^{(-)} + \gamma^{(0)}\right)B_{1,t}^{(min)} - \gamma^{(0)}B_{2,t} + \left(\gamma^{(+)} + \gamma^{(0)}\right)B_{3,t}^{(max)} =$$

$$= \left(\gamma^{(-)} + \gamma^{(+)}\right)B_{1,t}^{(min)} - \gamma^{(0)}B_{2,t} + \left(\gamma^{(+)} + \gamma^{(0)}\right)|B_{3t}|.$$

The graphs of ten trajectories of $W_t^{(2,-1,2)} = B_{1,t}^{(min)} + B_{2,t} + B_{3,t}^{(max)}, 0 \leq t \leq 10$, and $W^{(11,-1,2)} = 10B_{1,t}^{(min)} + B_{2,t} + B_{3,t}^{(max)}, 0 \leq t \leq 10$, are depicted in Figures B4 and B5.

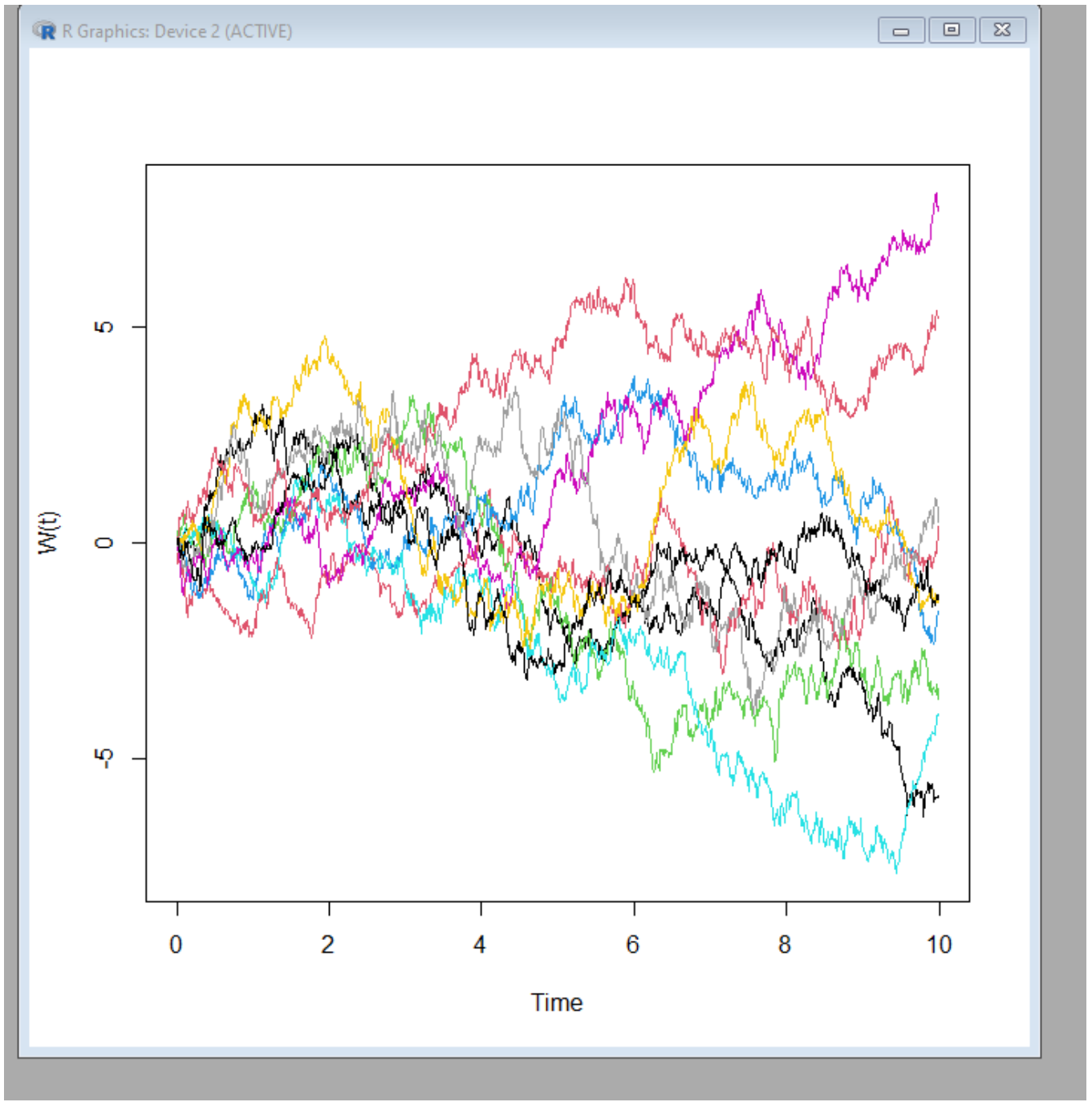

**Fig. B4** Ten trajectories of $W_t^{(2,-1,2)} = B_{1,t}^{(min)} + B_{2,t} + B_{3,t}^{(max)}, 0 \leq t \leq 10$.

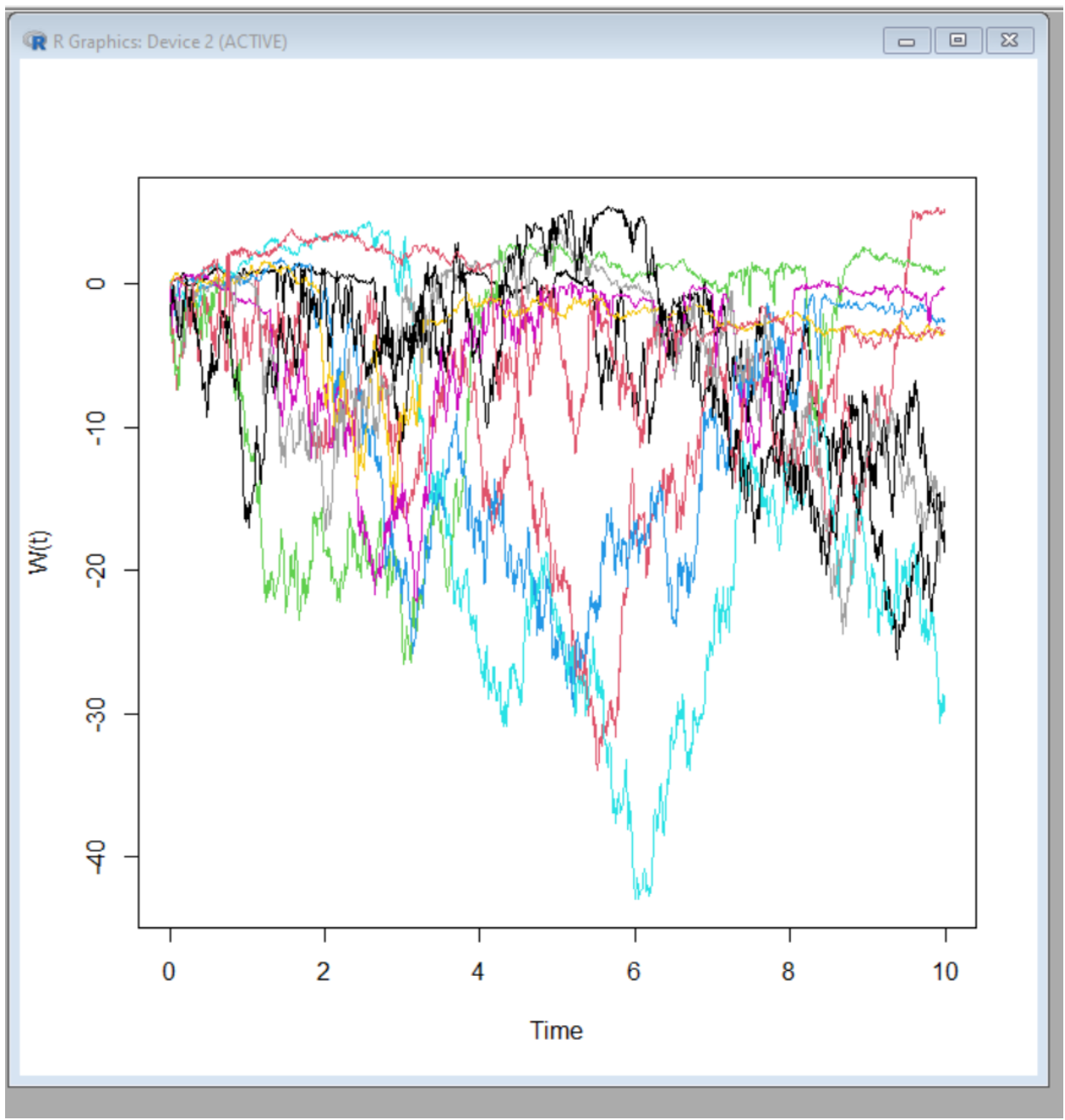

**Fig. B5** Ten trajectories of $W^{(11,-1,2)} = 10B_{1,t}^{(min)} + B_{2,t} + B_{3,t}^{(max)}, 0 \leq t \leq 10$.

The Itô-McKeen skew Brownian motion (SBM) $\mathbb{V}_t^{(\delta)}, \delta(-1,1)$, is defined as

$$\mathbb{V}_t^{(\delta)} = \sqrt{1 - \delta^2} B_{1,t} + \delta |B_{2,t}|,$$

where $B_{1,t}, t \geq 0$, and $B_{2,t}, t \geq 0$, are independent Brownian motions[35], and thus, it is a special case of the generalized skew Brownian motion.

Option pricing formulas based on GSBM and the corresponding **generalized skew random walk (GSRW)** will be of interest[36]. These formulas can be derived similarly to the formulas derived in [10, 74, 75]. The main link between discrete-time (based on the GSRW) and continuous-time option pricing (based on GSBM) will be determined by the CSYIP.

References

[1] Bachelier, L.: Th´eorie de la sp´eculation. Annales Scientifiques de l'Ecole Normale ´Sup´erieure 3(17), 21–86 (1900)

[2] Sullivan, E.J., Weithers, T.M.: Louis Bachelier: The father of modern option pricing theory. The Journal of Economic Education 22, 165–171 (1991) https: //doi.org/10.2307/1182421

[3] Courtault, J.M., Kabanov, Y., Bru, B., Cr´epel, P., Lebon, I., Marchand, A.L.: Louis Bachelier on the centenary of Th´eorie de la Sp´eculation. Mathematical Finance 10(3), 339–353 (2000) https://doi.org/10.1111/1467-9965.00098

[4] Davis, M., Etheridge, A.: Louis Bachelier's Theory of Speculations, The Origins of Modern Finance. Princeton University Press, Princeton, NJ (2006)

[5] Shiryaev, A.N.: Essentials of Stochastic Finance; Facts, Models, Theory. World Scientific Publishing Company, London (2003) 29

---

[35] The trajectories $\mathbb{V}_t^{(\delta)}(\omega), \omega \in \Omega$, of the Itô-McKeen skew Brownian motion (SBM) can be simulated using the alternative representation $\mathbb{V}_t^{(\delta)}(\omega) = \begin{cases} |B_t(\omega)|, w.p. \, \alpha \in (0,1), \\ -|B_t(\omega)|, w.p. \, 1-\alpha, \end{cases}$ where $B_t(\omega), \omega \in \Omega$, is a standard Brownian motion and $\alpha = \frac{1+\delta}{2}$.

[36] Corns and Satchell [75], Pasricha and He [76, 77], and Hu et al. [74] provide option pricing under skew Brownian motion and a skew random walk. The functional limit theorem (the invariance principle) for skew Brownian motion has a long history (see the references in [10, 74]) and is a particular case of the CSYIP, as shown in [59].

[6] Terakado, S.: On the option pricing formula based on the Bachelier model. Available at SSRN: https://ssrn.com/abstract=3428994 or http://dx.doi.org/10.2139/ ssrn.3428994 (2019)

[7] Brooks, R.E., Brooks, J.A.: Option valuation based on arithmetic Brownian motion: Justification and implementation issues. Available at SSRN: https://ssrn. com/abstract=2228288 or http://dx.doi.org/10.2139/ssrn.2228288 (2017)

[8] WPR: Countries with negative interest rates 2023. World Population Review. https://worldpopulationreview.com/country-rankings/ countries-with-negative-interest-rates (2023)

[9] Lauria, D., Lindquist, W.B., Mittnik, S., Rachev, S.T.: ESG-valued portfolio optimization and dynamic asset pricing. arXiv.org. http://arxiv.org/pdf/2206.02854 (2022)

[10] Hu, Y., Lindquist, W.B., Rachev, S.T.: ESG-valued discrete option pricing in complete markets. arXiv.org. http://arxiv.org/pdf/2209.06276 (2022)

[11] Hale, J.: New ESG rule for retirement plans: Plans should treat sustainability as any other relevant factor. Morningstar, Sustainability Matters. https://apple.news/AqWIMW9PsQTiBi79IoZFa9A (2023)

[12] Hu, Y., Shirvani, A., Lindquist, W..B., Fabozzi, F.J., Rachev, S.T.: Option pricing incorporating factor dynamics in complete markets. Journal of Risk and Financial Management 13(12), 321 (2020) https://doi.org/10.3390/jrfm13120321

[13] Hu, Y., Shirvani, A., Stoyanov, S., Kim, Y.-S., Fabozzi, F.J., Rachev, S.T.: Option pricing in markets with informed traders. International Journal of Theoretical and Applied Finance 23(13), 2050037 (2020)

[14] Hu, Y., Shirvani, A., Lindquist, W.B., Rachev, S.T., Fabozzi, F.J.: Market complete option valuation using a Jarrow-Rudd pricing tree with skewness and kurtosis. Journal of Economic Dynamics and Control 137(C) (2021). https: //www.sciencedirect.com/science/article/abs/pii/S0165188922000501

[15] Thomson, I.: Option pricing model: Comparing Louis Bachelier with Black Scholes Merton. Economics, Econometric Modeling: Derivatives eJournal. Available at SSRN: https://ssrn.com/abstract=2782719 or http://dx.doi.org/10.2139/ ssrn.2782719 (2016)

[16] Dimitroff, G., Fries, C.P., Lichtner, M., Rodi, N.: Lognormal vs normal volatilities and sensitivities in practice. Available at SSRN: https://ssrn.com/abstract= 2687742 or http://dx.doi.org/10.2139/ssrn.2687742 (2016)

[17] Duffie, D.: Dynamic Asset Pricing Theory. Princeton University Press, Princeton, NJ (2001) 30

[18] PIMCO: Negative interest rates. How do negative interest rates work? https://www.pimco.com/gbl/en/resources/education/ investing-in-a-negative-interest-rate-world (2016)

[19] S&P Global: Negative interest rates. https://www.spglobal.com/en/ research-insights/articles/negative-interest-rates (2016)

[20] Bech, M., Malkhozov, A.: How have central banks implemented negative policy rates? BIS Quarterly Review, 31–44 (2016). https://www.bis.org/publ/qtrpdf/ r qt1603e.pdf

[21] Lilley, A., Rogoff, K.S.: The case for implementing effective negative interest rate policy. Available at SSRN: https://ssrn.com/abstract=3427388 or http://dx.doi. org/10.2139/ssrn.3427388 (2019)

[22] Haksar, V., Kopp, E.: How can interest rates be negative? Central banks are starting to experiment with negative interest rates to stimulate their countries' economies. Finance & Development, 50–51 (2020). https://www.imf.org/ en/Publications/fandd/issues/2020/03/what-are-negative-interest-rates-basics

[23] Andersson, F.N.G., Jonung, L.: Lessons from the Swedish experience with negative central bank rates. Cato Journal 40(3) (2020) https://doi.org/10.36009/CJ. 40.3.2

[24] Claeys, G.: What are the effects of the ECB's negative interest rate policy? Policy Department for Economic, Scientific and Quality of Life Policies Directorate General for

Internal Policies, Monetary Dialogue Papers. https://www.europarl.
  uropa.eu/cmsdata/235691/02.%20BRUEGEL formatted.pdf (2021)

[25] Ulate, M., Lofton, O.: How do low and negative interest rates affect banks? FRBSF Economic Letter, Research from Federal Reserve Bank of San Francisco. https://www.frbsf.org/wp-content/uploads/sites/4/el2021-23.pdf (2021)

[26] Vaidya, D.: Spread meaning. WallStreet Mojo. https://www.wallstreetmojo.com/ spread/ (2023)

[27] Schaefer, M.P.: Pricing and hedging European options on futures spreads using the Bachelier spread option model. Paper presented at the NCR-134 Conference on Applied Commodity Price Analysis, Forecasting, and Market Risk Management, St. Louis, Missouri, 22–23 April 2002. https://legacy.farmdoc.illinois.edu/ nccc134/conf 2002/pdf/confp17-02.pdf (2002)

[28] Downey, L.: Spread option: Definition, examples, and strategies. Investopedia, Options and Derivatives, Advanced Concepts. https://www.investopedia.com/ terms/s/spreadoption (2022)

[29] Hayes, A.: Simple interest definition: Who benefits, with formula and example. 31 Investopedia, Personal Finance and Banking. https://www.investopedia.com/ terms/s/simple interest.asp (2023)

[30] Kenton, W.: Risk-free return calculations and examples. Investopedia Investing Basics. https://www.investopedia.com/terms/r/risk-freereturn.asp (2022)

[31] Rachev, S.T., Stoyanov, S.V., Fabozzi, F.: Financial markets with no riskless (safe) asset. International Journal of Theoretical and Applied Finance 20(8), 1–24 (2017)

[32] Karatzas, I., Shreve, E.: Brownian Motion and Stochastic Calculus. Springer Verlag, New York (1988)

[33] Melnikov, A., Wan, H.: On modifications of the Bachelier model. Annals of Finance 17, 187–214 (2021) https://doi.org/10.1007/s10436-020-00381-1

[34] Cox, J., Ross, S., Rubinstein, M.: Options pricing: A simplified approach. Journal of Financial Economics 7, 229–263 (1979)

[35] Jarrow, R.A., Rudd, A.: Option Pricing. Irwin, Homewood, IL (1983)

[36] Hull, J.: Options, Futures, and Other Derivatives, Eighth Edition. Pearson, Hoboken, NJ (2012)

[37] Shreve, S.S.: Stochastic Calculus for Finance I, The Binomial Asset Pricing Model. Springer, New York (2004)

[38] Shreve, S.S.: Stochastic Calculus for Finance II, Continuous-Time Finance. Springer, New York (2004)

[39] Dorfleitner, G.: Why the return notion matters. CORE World Scientific Pub Co Pte Lt. https://core.ac.uk/reader/11533729 (2003)

[40] Meucci, A.: Quant Nugget 2: Linear vs. compounded returns — Common pitfalls in portfolio management. GARP Risk Professional. Available at SSRN: https://ssrn.com/abstract=1586656 (2010)

[41] Hudson, R., Gregoriou, A.: Calculating and comparing security returns is harder than you think: A comparison between logarithmic and simple returns. Available at SSRN: https://ssrn.com/abstract=1549328 or http://dx.doi.org/10.2139/ssrn. 1549328 (2010)

[42] Jamison, M.: Why we use log returns for stock returns. Data Driven Investor. https://medium.datadriveninvestor.com/ why-we-use-log-returns-for-stock-returns-820cec4510ba (2022)

[43] Haugen, R.A.: Modern Investment Theory, Fourth Edition. Prentice-Hall Inc, Upper Saddle River, NJ (1997) 32

[44] Fama, E.F., French, K.R.: The capital asset pricing model: Theory and evidence. Journal of Economic Perspectives 18(3), 25–46 (2004)

[45] Lo, A.W.: The statistics of Sharpe ratios. Financial Analysts Journal 58(4), 36–52 (2002) https://doi.org/10.2469/faj.v58.n4.2453

[46] Miskolczi, P.: Note on simple and logarithmic return. CORE 11(1-2), 127–136 (2017). https://core.ac.uk/download/pdf/161062652.pdf

[47] Jarrow, R.A., Protter, P., Hasanjan, S.: No arbitrage without semimartingales. Ann. Appl. Probab. 19(2), 596–616 (2009)

[48] Delbaen, F., Schachermayer, W.: A general version of the fundamental theorem of asset pricing. Math. Ann. 300, 463–520 (1994)

[49] Delbaen, F., Schachermayer, W.: The fundamental theorem of asset pricing for unbounded stochastic processes. Math. Ann. 312, 215–250 (1998)

[50] Dzhaparidze, K., Zuijlen, M.C.A.: Introduction to option pricing in a securities market I: Binary models. CWI Quarterly 9(4), 319–355 (1996).
https://citeseerx.ist.psu.edu/document?repid=rep1&type=pdf&doi=
f2171136766ba573e7001a610a243988779328ec

[51] Cordero, F., Klein, I., Perez-Ostafe, L.: Binary markets under transaction costs. International Journal of Theoretical and Applied Finance 17(5), 1450030 (2014).
https://www.worldscientific.com/doi/10.1142/S0219024914500307

[52] Roll, R.: A simple implicit measure of the effective bid-ask spread in an efficient market. The Journal of Finance 39, 1127–1139 (1984)

[53] Hasbrouck, J.: Empirical Market Microstructure. Oxford University Press, New York (2007)

[54] Lauria, D., Hu, Y., Lindquist, W.B., Rachev, S.T.: Reconciling market microstructure and dynamic asset pricing theory. Manuscript in preparation (2023)

[55] Skorokhod, A.V.: Limit theorems for stochastic processes. Theory of Probability and its Applications 1, 261–290 (1956)

[56] Billingsley, P.: Convergence of Probability Measures, Second Edition. WileyInterscience, New York (1999)

[57] Davydov, Y., Rotar, V.: On a non-classical invariance principle. Statistics & Probability Letters 78, 2031–2038 (2008)

[58] Fama, E., French, K.: International tests of a five-factor asset pricing model. Journal of Financial Economics 123, 441–463 (2017) 33

[59] Cherny, A., Shiryaev, A., Yor, M.: Limit behavior of the "horizontal-vertical" random walk and some extensions of the Donsker-Prokhorov invariance principle. Theory of Probability and its Applications 47(3), 377–394 (2003)

[60] Jacod, J., Protter, P.: Discretization of Processes, Stochastic Modeling and Applied Probability, Vol. 67. Springer, Heidelberg (2012)

[61] Rachev, S., Mittnik, S.: Stable Paretian Models in Finance. John Wiley & Sons, Chichester, UK (2000)

[62] Cont, R.: Empirical properties of asset returns: s ✿ Stylized facts and statistical issues. Quantitative Finance 1, 223–236 (2001)

[63] Winegarden, W.: Environmental, social, and governance (ESG) investing: An evaluation of the evidence. Pacific Research Institute. https://www.sec.gov/ comments/climate-disclosure/cll12-8895812-241292.pdf (2019)

[64] OECD: ESG investing and climate transition: Market practices, issues and policy considerations. OECD Paris. https://www.oecd.org/finance/ ESG-investing-and-climatetransition-Market-practices-issues-and-policy-considerations. pdf (2021)

[65] Globerman, S.: ESG investing and asset returns, collected essays, ESG; Myths and realities. Fraser Institute. https://www.fraserinstitute.org/sites/default/ files/ESG-myths-realities-esg-investing-and-asset-returns 0.pdf (2022)

[66] Brock, T.: What is environmental, social, and governance (ESG) investing? Investopedia, Sustainable Investing, Socially Responsible Investing. https://www. investopedia.com/terms/e/environmental-social-and-governance-esg-criteria.asp (2023)

[67] Herstein, I.N., Milnor, J.: An axiomatic approach to measurable utility. Econometrica 21(2), 291–297 (1953)

[68] Smith, K.A.: Greenwashing and ESG: What you need to know. Forbes Advisor. https://www.forbes.com/advisor/investing/greenwashing-esg/ (2022)

[69] Bhagat, S.: An inconvenient truth about ESG investing. Investment Management, Harvard Business Review, Investment Management. https://hbr.org/2022/ 03/an-inconvenient-truth-about-esg-investing (2022)

[70] Webb, M.D.: ESG investing could end up being a classic mistake. MoneyWeek. https://moneyweek.com/investments/investment-strategy/esg-investing/ 604794/esg-investing-could-be-a-classic-mistake (2022)

[71] The Economist: The fundamental contradiction of ESG is being laid bare. Profit-seeking companies have too little incentive to save the 34 planet. The Economist, Leaders. https://www.economist.com/leaders/2022/09/ 29/the-fundamental-contradiction-of-esg-is-being-laid-bare (2022)

[72] Ardia, D., Boudt, K.: Implied expected returns and the choice of a mean-variance efficient portfolio proxy. Journal of Portfolio Management 41(4), 68–81 (2015)

[73] Chung, K.L., Williams, R.J.: Introduction to Stochastic Integration, Second Edition. Birkh¨auser, Boston (1990)

[74] Hu, Y., Lindquist, W.B., Rachev, S.T., Fabozzi, F.J.: Option pricing using a skew random walk pricing tree. arXiv preprint. https://arxiv.org/abs/2303.17014 (2023)

[75] Corns, T.R.A., Satchell, S.E.: Skew Brownian motion and pricing European options. The European Journal of Finance 13(6), 523–544 (2007) https://doi.org/10.1080/13518470701201488

[76] Pasricha, P., He, X.-J.: Skew-Brownian motion and pricing European exchange options. International Review of Financial Analysis 82©, 102120 (2022)

[77] Pasricha, P., He, X.: A simple European option pricing formula with a skew Brownian motion – ERRATUM. Probability in the Engineering and Informational Sciences 1(1) (2023) https://doi.org/10.1017/S0269964823000050